\documentclass[lettersize,journal]{IEEEtran}
\usepackage{amsmath,amsfonts}
\usepackage{algorithmic}
\renewcommand{\algorithmicrequire}{ \textbf{Input:}}

\usepackage{array}
\usepackage[caption=false,font=normalsize,labelfont=sf,textfont=sf]{subfig}
\usepackage{graphicx}
\def\BibTeX{{\rm B\kern-.05em{\sc i\kern-.025em b}\kern-.08em
    T\kern-.1667em\lower.7ex\hbox{E}\kern-.125emX}}
\usepackage{balance}
\usepackage{subfiles}
\usepackage{blindtext}
\usepackage{setspace}
\usepackage{mathrsfs}
\usepackage{algorithm}
\usepackage[T1]{fontenc}
\usepackage{booktabs}
\usepackage{xurl}
\usepackage{cite}
\usepackage{color}
\usepackage{multirow}
\usepackage{afterpage}
\usepackage{makecell}
\usepackage{colortbl} 
\usepackage{arydshln} 
\definecolor{myblue}{RGB}{0, 0, 0}
\definecolor{blue2}{RGB}{0, 0, 0}

\begin{document}
\title{Reducing Traffic Wastage in Video Streaming via Bandwidth-Efficient Bitrate Adaptation}

\author{Hairong~Su$^{\dag}$,
         Shibo~Wang$^{\dag}$,
         Shusen~Yang,
         Tianchi~Huang,
         Xuebin~Ren
\thanks{$^{\dag}$Co-first authors.}
\thanks{This work was supported in part by the National Key Research and Development Program of China under Grants 2021YFB2401300, 2022YFA1004100, and 2020YFA0713900; in part by the National Natural Science Foundation of China under Grants 61772410, 61802298, 62172329, U1811461, U21A6005, and 11690011.
\textit{(Corresponding author: Shusen Yang.)}}
\thanks{Hairong Su, Shibo Wang, Xuebin Ren are with the National Engineering Laboratory for Big Data Analytics, Xi'an Jiaotong University.}
\thanks{Shusen Yang is with the National Engineering Laboratory for Big Data Analytics, and the Ministry of Education Key Laboratory for Intelligent Networks and Network Security, Xi'an Jiaotong University.}
\thanks{Tianchi Huang is with the R\&D department, Sony.}}

\markboth{IEEE TRANSACTIONS ON MOBILE COMPUTING}%
{Shell \MakeLowercase{\textit{et al.}}: Bare Demo of IEEEtran.cls for Computer Society Journals}

\maketitle

\begin{abstract}
Bitrate adaptation (also known as ABR) is a crucial technique to improve the quality of experience (QoE) for video streaming applications.
However, existing ABR algorithms suffer from severe traffic wastage, which refers to the traffic cost of downloading the video segments that users do not finally consume, for example, due to early departure or video skipping.
In this paper, we carefully formulate the dynamics of buffered data volume (BDV), a strongly correlated indicator of traffic wastage, which, to the best of our knowledge, is the first time to rigorously clarify the effect of downloading plans on potential wastage.
To reduce wastage while keeping a high QoE, we present a bandwidth-efficient bitrate adaptation algorithm (named BE-ABR), achieving consistently low BDV without distinct QoE losses.
Specifically, we design a precise, time-aware transmission delay prediction model over the Transformer architecture, and develop a fine-grained buffer control scheme.
Through extensive experiments conducted on emulated and real network environments including WiFi, 4G, and 5G, we demonstrate that BE-ABR performs well in both QoE and bandwidth savings, enabling a 60.87\% wastage reduction and a comparable, or even better, QoE, compared to the state-of-the-art methods.
\end{abstract}


\section{Introduction}
\IEEEPARstart{V}{ideo} streaming has experienced tremendous growth in popularity over the last few decades, driven by advances in network technology \cite{5GVoD, Lever5G} and the rise of various video streaming services \cite{Netflix, Youtube, Tiktok, Zoom}.
In 2022, the global video streaming market was valued at USD 69.89 billion, with an expected compound annual growth rate (CAGR) of 20.50\% through 2030 \cite{VANTAGE}.
With video content currently accounting for over 65\% of total internet traffic \cite{Gutelle}, video streaming has become the dominant application on the Internet.

Bitrate adaptation, also known as ABR, plays a fundamental role in improving the QoE performance of video streaming. 
However, existing ABR algorithms unfortunately introduce considerable traffic wastage \cite{finamore2011youtube,li2023dashlet,kuaishou2021}, which refers to the bandwidth consumption that fails to end up with actual viewing by users. 
In fact, 25.2\%$\sim$51.7\% of downloaded video data was wasted according to the measurements of mainstream ABR algorithms \cite{zhang2021post}.

Excessive traffic wastage of video streaming adversely affects both consumers and content providers.
First, it may lead to heightened customer attrition and diminished usage frequency.
Mobile traffic remains costly today; \textcolor{myblue}{for instance, China Telecom \cite{ChinaTelecom2023} charges 10 RMB/GB for cellular traffic\footnote{\textcolor{myblue}{https://www.189.cn/bj/liuliang/}}.}
Hidden traffic wastage aggravates the financial burden of accessing video services via mobile networks \cite{HMDDSU2021}.
Second, traffic wastage significantly raises operational expenses (OPEX) for media-streaming companies.
It is reported that bandwidth costs are anticipated to become a major source of OPEX for vehicular media operators owing to heavy daily data consumption \cite{hong2018cost}.

Most traffic wastage is caused by the concurrence of sustaining \textit{large} buffer and users' \textit{erratic} behaviors \cite{huangbuffer}.
Current ABR agents \cite{yin2015control,huang2014buffer,yan2020learning} tend to preserve large buffer occupancy (typically, \textgreater 10 \textit{sec}), that is, prefetching superabundant content ahead of watching.
If the user abruptly exits or skips a large portion of the video, the content stored in the buffer would be completely discarded and wasted.
Such irregular viewing behaviors are quite prevalent; in fact, fewer than 10\% of users watch a video from beginning to end without skipping any parts \cite{conviva}. 

Despite the unpredictability of user behavior, traffic wastage can be mitigated by maintaining low buffer occupancy during streaming.
However, this presents challenges across several aspects.
First, pursuing low buffer occupancy actually refers to reducing the \textit{buffered data volume} (BDV), rather than the commonly referenced buffered video time (BVT) \cite{yin2015control}.
BDV and BVT are two metrics employed to assess the buffer state, signifying the amount and duration of video content stored within the buffer, respectively.
Although BVT at the moment a user leaves or skips can provide some indication of traffic wastage, BDV serves as a more direct measure.
Nonetheless, the impact of different download configurations on BDV, i.e., BDV dynamics, remains unclear.

Second, attaining high QoE performance given a limited buffer is non-trivial.
The reduced buffer occupancy imposes greater demands on the accuracy of throughput prediction, especially in the context of volatile mobile networks.
An inaccurate network predictor may lead to frequent playback rebuffering, as the buffer might not contain enough video segments to compensate for errors in throughput prediction. 
Consequently, the user's viewing experience can be significantly degraded.

Moreover, simply relying on bitrate selection is insufficient for precisely controlling the buffer to the desired level.
In most cases, none of the available bitrates closely matches the optimal rate decision made by the ABR controller.
This is because bandwidth is infinitely divisible, while the set of bitrate options tends to be quite finite ($<$5 options for most content providers) for reducing the storage overhead.
The discrepancy could lead to imprecise buffer control, particularly when the bandwidth far exceeds the maximum bitrate.
Therefore, it is necessary to introduce rational plannings of inter-chunk waiting time\footnote{The inter-chunk waiting time refers to the time gap between the completion of a chunk's fetching and the beginning of the subsequent chunk's fetching.} to gain more proactive and granular control over playback buffer; however, \textcolor{myblue}{in most ABR algorithms~\cite{yin2015control,yan2020learning,spiteri2020bola,huang2014buffer,mao2017neural}}, downloading is only paused when the buffer reaches its predetermined upper limit (e.g., 20 \textit{sec}).

To address the above challenges, we first model the dynamics of buffered data volume and formulate the wastage-aware rate adaptation into a stochastic optimization problem.
Then we introduce BE-ABR, a bandwidth-efficient bitrate adaptation algorithm, designed to minimize traffic wastage while maintaining high QoE performance. 
Specifically, we devise a Transformer-based, time-aware transmission delay predictor (called $\mathrm{T^{3}P}$) to forecast the transmission duration of each video chunk.
With a novel network architecture,
$\mathrm{T^{3}P}$ can effectively capture the evolution trends of irregular time-series network data, achieving precise predictions.

Based on relatively precise network predictions, we propose a rate adaptation algorithm that incorporates fine-grained buffer control to achieve both high-quality and low-cost streaming.
The quality level and the inter-chunk waiting time are jointly regulated to precisely maintain the buffer at an optimal level, balancing the need for smooth playback and efficient bandwidth utilization.
Furthermore, we present an adaptive weighting scheme based on network fluctuations and a QoE-constrained optimal configuration searching algorithm to enhance QoE performance robustness and maintain manageable QoE loss.

We implemented a DASH-based video streaming testbed and conducted a variety of experiments on real-world networks containing commodity WiFi and cellular networks in the wild.
Results derived from over 500 hours of streaming demonstrate that BE-ABR's overall performance is well ahead of several benchmarks.
In terms of the QoE metric, BE-ABR outperforms the best existing scheme in WiFi environments and matches it equally in cellular environments.
Furthermore, BE-ABR exhibits the lowest traffic waste in all evaluated scenarios, reducing it by 35.7\%$\sim$67.2\%.
Additionally, we conducted a component-wise study to investigate the individual contributions of key components in BE-ABR.
\textcolor{myblue}{
We also ran a user study to evaluate the performance under real-world viewing conditions.
Moreover, we evaluated BE-ABR under high-definition video configurations using numerous real-world traces (4G, 5G), further demonstrating its robustness in a variety of environments.}
Lastly, we evaluated $\mathrm{T^{3}P}$ separately, establishing its predictive superiority against other network prediction algorithms.
	
In summary, our key contributions are as follows:
\begin{enumerate}
    \item
    \textcolor{myblue}{
    We formulate the dynamics of buffered \textit{data volume} and develop an explicit mathematical model for the wastage-aware adaptive streaming problem,} laying the groundwork for wastage-controlled ABR design.
    \item
    We develop a high-accuracy transmission delay prediction model over a new time-aware Transformer deep neural network (DNN) architecture.
    \item 
    We propose a bandwidth-efficient bitrate adaptation algorithm (called BE-ABR), achieving consistently low buffer occupancy with \textit{guaranteed} QoE management.
    \item
    Through extensive experiments on emulated and real-world wireless networks, we demonstrate that BE-ABR significantly reduces video traffic wastage (by 60.87\%) while preserving high viewer QoE.
\end{enumerate}

The remainder of this paper is organized as follows.
In section \uppercase\expandafter{\romannumeral2}, we review related work.
In section \uppercase\expandafter{\romannumeral3}, we describe the video streaming system model, formulating the dynamics of buffered data volume.
In section \uppercase\expandafter{\romannumeral4}, we state the optimization problem for both QoE and traffic wastage.
In section \uppercase\expandafter{\romannumeral5}, we present the design of BE-ABR.
In section \uppercase\expandafter{\romannumeral6}, we evaluate the performance of BE-ABR through extensive experiments.
In section \uppercase\expandafter{\romannumeral7}, 
we discuss the limitation and future work.
In section \uppercase\expandafter{\romannumeral8},
we conclude the study.

\section{Related Work}
\subsection{Adaptive Bitrate Streaming}
The field of ABR has been studied for many years.
Liu \textit{et al.} \cite{liu2011rate} develop a throughput prediction-based approach that determines bitrate switching operations according to a smoothed HTTP throughput measured by segment fetch time.
Huang \textit{et al.} \cite{huang2014buffer} introduce a buffer-based approach (BBA) for rate adaptation problems, arguing that throughput prediction is unnecessary during steady-state phases and that video rate could be selected directly as a function of current buffer occupancy.
Yin \textit{et al.} \cite{yin2015control} propose MPC, a hybrid throughput-buffer-based bitrate adaptive algorithm that formulates the bitrate adaptation as a stochastic optimal control problem and design a model predictive control approach to maximize user QoE. 
Mao \textit{et al.} \cite{mao2017neural} suggest a learning-based approach, Pensieve, to generate ABR algorithms through reinforcement learning, training a neural network model to output the control policy for bitrate adaptation based on historical observations.

\textcolor{myblue}{As the performance of QoE has reached a certain level, recent years have seen a growing concern regarding the issue of traffic wastage.
Chen \textit{et al.} \cite{chen2015smart} and Huang \textit{et al.} \cite{huang2020online} curb the data waste in multiuser streaming systems through the proposal of wastage-aware resource allocation strategies.
Li \textit{et.al} \cite{li2022alfie, li2023dashlet, qian2022dam, zhou2022pdas} mitigate bandwidth wastage in the realm of short video services by optimizing preloading strategies based on predictions of swipe behavior.
However, effective solutions are still lacking for the wastage brought by inappropriate ABR algorithms in video-on-demand services.}
Zhang \textit{et al.} \cite{zhang2021post} propose PSWA, a broad framework that transforms existing ABR algorithms into wastage-aware versions.
However, its generic structure may impede optimal performance.
In contrast to previous work, BE-ABR presented in this paper provides a high-quality and cost-effective rate adaptation policy by accurately maintaining the buffer at an optimal level that balances QoE and wastage.

\subsection{Network Condition Prediction}\label{subsec:RW_TTP}
Network condition prediction is a fundamental issue in video streaming systems, aiming to predict future network conditions with historical network information.
Researchers have produced a literature of statistical or machine-learning methods to predict network throughput.

In \cite{jiang2012improving} and \cite{yin2015control}, the authors use harmonic mean values over the past several chunks to predict TCP throughput when downloading the next chunk.
A hidden Markov model (HMM) is proposed in CS2P \cite{sun2016cs2p} modeling the stateful evolution of intra-session throughput.
The researchers of \cite{wei2018throughput,mei2019realtime,raca2020leveraging} apply deep learning models such as the recurrent neural network (RNN) model, long short-term memory (LSTM) model and support vector machines (SVM) model to make real-time bandwidth predictions.
These works assume that future bandwidth is unaffected by download strategies, ignoring the impact of the requested data volume on bandwidth allocation \cite{bartulovic2017biases,zhang2017modeling}.

Differing from above works, Fugu \cite{yan2020learning} directly predicts the transmission time with a fully connected network.
However, the model employed in Fugu fails to effectively capture intricate bandwidth patterns due to its limited temporal information processing and network features extraction ability.
In this paper, we thoroughly explore network characteristics and propose a novel, time-aware, Transformer-based neural network architecture.
This innovative network structure enables high-precision predictions of transmission delay for each chunk.

\begin{figure}[tbp]
    \centerline{\includegraphics[width=2.8in]{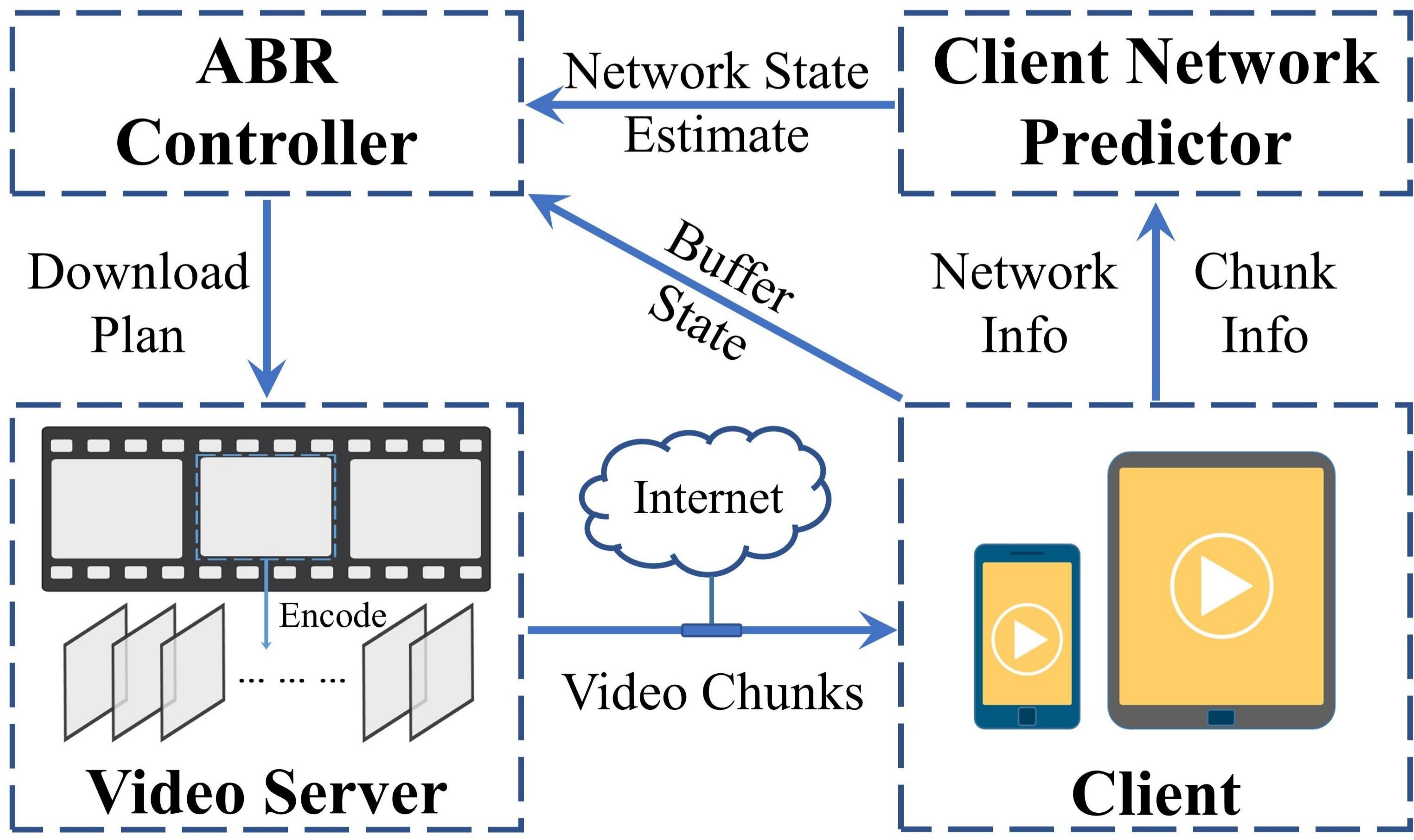}}
    \caption{An overview of DASH-based video streaming system.}
    \label{SystemOverview}
\end{figure}

\section{System Model}
As shown in Fig.~\ref{SystemOverview}, we consider a DASH-based video streaming system.
Videos are preprocessed and stored on the server, encoded into multiple chunks with varying quality levels before transmission.
During the streaming process, the client sequentially requests video chunks with specified bitrate levels from the server, following the download plan determined by the ABR controller.
Subsequently, we will expound upon the mathematical model of the DASH-based streaming, providing a formulaic description of buffered data volume dynamics.

\begin{table}[!tbp]
\footnotesize
\centering
\label{notation}
\arrayrulecolor{myblue}
\color{myblue}
\caption{Notation Used}
\begin{tabular}{c|m{6.65cm}}
\hline
\textbf{Notation} & \textbf{Description} \\
\hline
$K$ & The total number of chunks \\
\hline
$L$ & The duration of each chunk \\
\hline
$\mathcal{R}$ & The available bitrate levels \\
\hline
$L_{max}$ & The physical upper limit of the buffer \\
\hline
$d_k(R_k)$ & The size of the chunk $k$ at bitrate $R_k$ \\
\hline
$q(R_k)$ & The quality of chunk $k$ at bitrate $R_k$ \\
\hline
$D_k(R_k)$ & The download time of chunk $k$ at bitrate $R_k$ \\
\hline
$t_k$ & The time when the download of chunk $k$ starts \\
\hline
$t_k^{\prime}$ & The time when the download of chunk $k$ completes \\
\hline
$\Delta t_k$ & The waiting time before the download of chunk $k+1$ \\
\hline
$L(t)$ & The length of the video cached in the buffer at time $t$\\
\hline
$S(t)$ & The volume of the video cached in the buffer at time $t$ \\
\hline
$V(t)$ & The media time of the video being played at time $t$ \\
\hline
$M(v)$ & The download bitrate for the $v$-th second of the video \\
\hline
$H(t_1, t_2)$ & The volume of video consumed by user viewing over the interval $(t_1, t_2)$ \\ 
\hline
\end{tabular}
\end{table}

\textbf{On the server side:} The video is divided into $K$ equal-length chunks of duration $L$, encoded at a finite list of bitrate levels $\mathcal{R}$.
\textcolor{myblue}{
The size of chunk $k$ encoded at bitrate $R_k$ is represented by $d_k(R_k)$.
This information can be pre-acquired by the client side by parsing the manifest file before transmission.}
A non-decreasing function $q(\cdot): \mathcal{R} \rightarrow R_+$, is implemented to map bitrates to quality, rendering the quality of chunk $k$ at bitrate $R_k$ as $q(R_k)$.

\textbf{On the client side:} The downloaded but unviewed video is stored in the client-side playback buffer. 
Let $L(t)$ represent the remaining video time in the buffer at time $t$. 
The playback buffer has a physical upper limit $L_{max}$; once reached, the server cannot transmit additional segments. 
Naturally, $L_{max}$ should be greater than the chunk length $L$ to prevent buffer overflow upon receiving a new chunk.

\textbf{Downloading process:}
At time $t_k$, the client starts to download chunk $k$ from the server. 
The download time, $D_k(R_k)$, is dictated by the chunk size $d_k(R_k)$ and the future bandwidth $B(t)$ during the downloading process of chunk $k$.
We introduce $\mathrm{T^{3}P}$ to provide an accurate prediction of the download time $D_k(R_k)$ and discuss it further in Section \ref{subsec:t3p}.

Upon completing the download of chunk $k$, the video player pauses for $\Delta t_k$ before downloading chunk $k+1$.
The duration of this pause, $\Delta t_k$, can be set within a certain range.
\textcolor{myblue}{On one hand, if the buffer exceeds $L_{max}$ after the addition of chunk $k$, the player needs to wait until the buffer falls below the threshold:}
\begin{equation}
	\Delta t_k \geq \left(\left(L_k - D_k(R_k)\right)_{+} + L - L_{max}\right)_{+}.
\end{equation}
\textcolor{myblue}{Here, $(x)_+=max\{x,0\}$.}
On the other hand, it should not result in the occurrence of rebuffer events for the sake of QoE performance:
\begin{equation}
	\Delta t_k \leq \left(L_k - D_k(R_k)\right)_{+} + L.
\end{equation}
Here we denote $L(t_k)$ as $L_k$ for brevity.
To precisely maintain the buffer at the optimal level, we plan for the waiting time $\Delta t_k$ and elaborate on it in Section \ref{subsec:abr}.
Then the time start to download chunk $k+1$ could be calculated as:
\begin{equation}
	t_{k+1} = t_k + D_k(R_k) + \Delta t_k.
\end{equation}

At time $t_{k+1}$, considering the download of chunk $k$ and the playback consumption caused by user viewing, $L_{k+1}$ should be expressed as:
\begin{equation}
	L_{k+1} = \left(L_k - D_k(R_k)\right)_+ + L - \Delta t_k.
\end{equation}

\textbf{Dynamics of buffered data volume:}
\textcolor{myblue}{
The buffered data volume, denoted as $S(t)$, is a critical metric in wastage-aware adaptive streaming, signifying the volume of video data cached in the buffer at time $t$.
It equates to the amount of data waste that would occur if the user discontinues viewing at time $t$.
The following provides expressions for the trajectory of $S(t)$, illustrating how it is affected by download configurations, and paves the way for the formulation of the wastage-aware adaptive streaming problem in the subsequent section.}

\textcolor{myblue}{$S(t)$ evolves over time as influenced by two primary factors. One is the consumption caused by user viewing, and the other is the supplement from the video chunks being downloaded into the buffer.
We denote the volume of video consumed by user viewing during a time period $(t_1,t_2)$ as $H(t_1, t_2)$.
Its calculation formula is as follows:}
\begin{equation}
{\color{myblue} H(t_1, t_2) = \int_{V(t_1)}^{V(t_2)} M(v)dv.}
\end{equation}
\textcolor{myblue}{
In this equation, $M(v)$ represents the download bitrate for the $v$-th second of the video.
And $V(t)$ represents the media time of the video being played at the natural time $t$.
Here, natural time $t$ reflects the actual elapsed time from when the user starts watching the video, while the media time $V(t)$, refers to the point in time in the video at natural time $t$.
It means the user has watched up to the $V(t)$-th second of the video at time $t$.
If the expressions for $M(v)$ and $V(t)$ are provided, we can compute the buffer consumption $H(t_1, t_2)$ for any given time period.
It's easy to understand that $M(v)$ corresponds to the download bitrate of the video:}
\begin{equation}
	M(v) = R_k, \quad v\in \left(\left(k-1\right)L, kL\right].
\end{equation}

\textcolor{myblue}{
Next, we provide the expression for $V(t)$ in the interval $t\in (t_k, t_{k+1})$.
At time $t_k$, given that $(k-1)L$ seconds of the video have been downloaded and $L_k$ seconds of unviewed content remain in the buffer, $V(t_k)$ can be inferred from $L_k$ as follows:}
\begin{equation}
V(t_k) = (k-1)L - L_k.
\end{equation}
\textcolor{myblue}{Considering the potential for rebuffering events, we need to explore two cases: a) $D_k(R_k) \leq L_k$; b) $D_k(R_k) \textgreater L_k$.}

\textcolor{myblue}{For case a), where $D_k(R_k) \leq L_k$, rebuffering events will not occur and the evolution of $V(t)$ is described by:}
\begin{equation}
    {\color{myblue} V(t) = V(t_k) + t - t_k, \hspace{0.15cm} t\in \left(t_k, t_{k+1}\right].}
\end{equation}
\textcolor{myblue}{
In this scenario, since there are no rebuffering events, the user's viewing progress will proceed in sync with the progression of actual time.}

\textcolor{myblue}{For case b), where $D_k(R_k) \textgreater L_k$, rebuffering events will occur, causing the viewer's progress $V(t)$ to evolve differently:}
\begin{equation}
\color{myblue}{
    \begin{aligned}
        V(t)=
        &\begin{cases}
        V(t_k) + t-t_k, \hspace{0.15cm} t\in \left(t_k, t_k + L_k\right]. \vspace{1ex}\\
        V(t_k) + L_k, \hspace{0.15cm} t\in \left(t_k + L_k, t_k^{\prime}\right]. \vspace{1ex}\\
        V(t_k^{\prime}) + t- t_k^{\prime}, \hspace{0.15cm} t\in \left(t_k^{\prime}, t_{k+1}\right].
        \end{cases}
    \end{aligned}}
\end{equation}
\textcolor{myblue}{The fundamental difference between case b) and case a) is that during rebuffering events, the viewer's progress will freeze at the current timestamp. 
Here, $t_k^{\prime}$ represents the moment when chunk $k$ is completely downloaded, expressed as:}
\begin{equation}
{\color{myblue} t_k^{\prime} = t_k + D_k(R_k).}
\end{equation}

\textcolor{myblue}{
Having modeled the key variables, we can then formulate expressions for $S(t)$ within the time interval $(t_k, t_{k+1}]$.}
\textcolor{myblue}{
We segment this interval into two stages contingent on whether chunk $k$ has been fully downloaded: 1) $(t_k, t_k^{\prime}]$; 2) $(t_k^{\prime}, t_{k+1}]$.}

First, we analyze the changes in $S(t)$ during $(t_k, t_k^{\prime}]$, where the download of chunk $k$ is not yet complete.
In this stage, segments of chunk $k$ are successively downloaded into the buffer, while there is viewing consumption in the meantime.
Hence, $S(t)$ evolves according to the following equation:
\begin{equation}
    {\color{myblue} S(t) = S(t_k) - H(t_k, t) + d_k(R_k) \times \frac{t-t_k}{D_k(R_k)}, \hspace{0.1cm} t \in \left(t_k, t_k^{\prime} \right].}
\end{equation}
In the equation above, the second term represents the buffer consumption caused by user viewing, while the third term signifies the buffer replenishment resulting from downloading.
Note that, when calculating downloading replenishment, we assume the bandwidth stays consistent over a brief time span.

Then we will proceed to model the dynamics of $S(t)$ during $(t_k^{\prime}, t_{k+1}]$.
In this stage, the buffer only experiences consumption without any replenishment because the download of chunk $k$ has been completed.
Thus, $S(t)$ evolves according to:
\begin{equation}
\color{myblue}{
S(t) = S(t_k^{\prime}) - H(t_k^{\prime}, t), \hspace{0.15cm} t\in(t_k^{\prime}, t_{k+1}].
}
\end{equation}
\textcolor{myblue}{We've now derived the trajectory expressions for $S(t)$, which equals to the size of data wastage should the user leave at a certain future time $t$. With these in hand, we can proceed to the problem formulation section.}

\section{Problem Formulation}
The objective of this study is to achieve both high-quality and cost-effective video streaming, which entails preserving the QoE performance while reducing unnecessary network traffic.
Typically, traffic wastage occurs when a user abruptly quits or switches to a different playback point.
This paper focuses exclusively on the impact of early departure behavior.
If the user skips a large block of video content, it could be treated in the same way as early departure behavior because of comparable effects.
And if the user skips a small segment of the video content, it could be overlooked because it has a minor impact on streaming.

The performance of streaming should be discussed in two dimensions: QoE scores and traffic wastage.
To assess QoE performance, we consider three widely used measurements: video quality, which measures the average video quality perceived by the user; bitrate variations, which quantify the extent of quality changes between successive chunks; and rebuffering ratio, which is the ratio of total rebuffering time to total playback time.

Assuming that the user leaves at time $t_{0}$, we can denote the number of fully viewed chunks as $k_1(t_{0})$ and the number of already downloaded chunks as $k_2(t_{0})$.
The QoE scores obtained by the user leaving at time $t_{0}$ can then be calculated as:
\begin{equation}
	\begin{aligned}
		QoE(t_{0}) = &\alpha_1\times\frac{1}{k_1(t_{0})}\sum\limits_{k=1}^{k_1(t_{0})}QUA_k - \alpha_2\times\frac{1}{t_{0}}\sum\limits_{k=1}^{k_2(t_{0})}REB_k \\ &-  \alpha_3\times\frac{1}{k_1(t_{0})-1}\sum\limits_{k=2}^{k_1(t_{0})}VAR_k,
	\end{aligned}
 \label{eq:9}
\end{equation}
where, 
\begin{equation}
\begin{cases}
	&QUA_k = q(R_k),\\
	&VAR_k = |q(R_k) - q(R_{k-1})|,\\ 
	&REB_k = (D_k(R_k) - L_k)_+.
\end{cases}
\end{equation}
Here, we respectively use $QUA_k, REB_k, VAR_k$ to represent the video quality perceived by the user of chunk $k$, the rebuffering time resulting from downloading chunk $k$, and the bitrate variations between chunk $k$ and chunk $k-1$.
Additionally, $\alpha_1, \alpha_2, \alpha_3$ are non-negative weighting parameters of the three QoE metrics.
It is crucial to note that only the chunks consumed by the client should be included in the calculation, rather than all the downloaded chunks.
For instance, if the client leaves before the playback of chunk $k$, the quality of chunk $k$ should not be considered, even though it has already been downloaded.

The traffic wastage that ensues when a user departs equates to the volume of buffered data remaining at the point the user exits. 
Therefore, the traffic wastage resulting from a user's exit at time $t_{0}$ can be expressed as:
\begin{equation}
	WAS(t_{0}) = S(t_{0}).
\end{equation}
The magnitude of $WAS(t_{0})$ is contingent upon the discrepancies in volume between the consumed video and the downloaded video at time $t_{0}$.

Now we can define the reward for the user leaving at time $t_{0}$ as:
\begin{equation}
	REWARD(t_{0}) = QoE(t_{0}) - \beta \times WAS(t_{0}).
\end{equation}
Here $\beta$ is a non-negative weight coefficient adjusting the proportions of the above terms in the reward function.
To ensure a high-quality user experience and save bandwidth resources at the same time, the optimization problem we need to solve is:
\begin{equation}
\begin{aligned}
    \mathbf{Problem\ 1:} \qquad & max \quad REWARD(t_{0}),\hspace{2cm}\\
    &s.t. \quad \text{Constraints (1) to (10)}.
\end{aligned}
\end{equation}

Ideally, we want to maximize the $REWARD(t_{0})$ to achieve heightened user satisfaction and lower traffic wastage.
Nonetheless, solving \textit{Problem 1} requires the prediction of download times for all future $k_2(t_{0})$ chunks at each bitrate level, which is impractical due to the complex and ever-changing network environments.
Moreover, obtaining an oracle that reports the departure time $t_{0}$ for each client before the start of streaming is unachievable.
First, viewing behavior exhibits individual diversities, with minimal similarity among different users \cite{li2023dashlet,zhang2020apl}.
Second, departure behaviors vary across videos with different types and durations.
Third, the quality of current video services can also influence user behaviors.
For example, frequent rebuffering events can significantly diminish user engagement, potentially resulting in more frequent departures.
To this end, we develop BE-ABR in this study and introduce it in the following section.

\section{BE-ABR Algorithm}
In this section, we clarify the core principles underlying the design of BE-ABR, which integrates a Transformer-based, time-aware transmission delay predictor ($\mathrm{T^{3}P}$) with buffer-controlled rate adaptation.
Initially, we decompose the global optimization problem into a series of local optimization problems to approach the global optimal solution.
Subsequently, we provide a concise overview of the BE-ABR workflow.
Following this, we introduce the innovative network structure of $\mathrm{T^{3}P}$, designed to precisely predict the transmission time of each chunk.
Finally, we discuss the rate adaptation algorithm with fine-grained buffer control, designed to effectively  manage the buffer at an optimal level that balances QoE and wastage.

\subsection{Problem Transformation}
As previously stated, video streaming systems exhibit considerable uncertainty due to the variability in both bandwidth evolution and user behaviors, complicating global optimization.
To address this challenge, we divide the global problem into a sequence of local optimization problems.
This strategy enables us to adjust the control strategy at each time step based on the latest feedback and information, leading to more stable control and efficient optimization.
Instead of considering the entire time range (i.e., from the present moment until the user stops watching), we only need to focus on the variables and constraints for the immediate future (i.e., subsequent $N$ chunks).
This is feasible since near-future bandwidth can be predicted to some extent using historical data, given that network connections tend to remain relatively stable and continuous within a certain time period \cite{zhang2001constancy}.

At each time step $t_k$, we plan a control trajectory for the subsequent $N$ chunks to optimize their expected total reward $REWARD_{k}^{k+N}$.
The QoE scores for chunks $k$ to $k+N$ are computed as follows:
\begin{equation}
  \begin{aligned}
    QoE_k^{k+N} = &\alpha_1 \tiny{\times} \frac{1}{N} \sum_{i=k}^{k+N}QUA_i - 
    \alpha_2 \tiny{\times} \frac{1}{t_{k+N}-t_k} \sum_{i=k}^{k+N}REB_i \\ &- \alpha_3 \tiny{\times} \frac{1}{N} \sum_{i=k}^{k+N}VAR_i.
  \end{aligned}
\end{equation}

Although accurately forecasting either the timing of a user's departure or the likelihood of their leaving at a specific future point is infeasible, maintaining a consistently small buffer would naturally contribute to a reduction in traffic wastage.
Thus, $WAS_{k}^{k+N}$ could be defined as the average buffered data volume during the download of the next $N$ chunks:
\begin{equation}
	WAS_{k}^{k+N} = \frac{1}{t_{k+N}-t_k}\sum_{i=k}^{k+N}\int_{t_i}^{t_{i+1}}S(t)dt.
\end{equation}

Given all this, solving the original \textit{Problem 1} is transformed into maximizing \textit{Problem 2} at each time step $t_k$:
\begin{equation}
	\begin{aligned}
		\mathbf{Problem\ 2:} \quad \quad &
        max \quad REWARD_k^{k+N} \hspace{2cm}\\
		= &QoE_k^{k+N} - \beta \times WAS_k^{k+N},\\
        s.t. \quad &\text{Constraints (1) to (10)}.
	\end{aligned}	
 \label{eq:16}
\end{equation}

\begin{algorithm}[!t]
\caption{BE-ABR's Workflow.}
\renewcommand{\algorithmicrequire}{ \textbf{Comments:}}
\begin{algorithmic}
\fontsize{9.5}{12}\selectfont
\REQUIRE
\STATE $\text{AdapWght}$: Network fluctuations based adaptive weighting.
\STATE $\text{Srch}$: QoE-constraint optimal configuration searching.
\item[] $\text{Opts}:\{(R_k, \cdots, R_{k+N},\Delta t_k,\cdots, \Delta t_{k+N})\big|R_i \in \mathcal{R}, \Delta t_i \in \mathcal{T}, i = k,\cdots, k+N\}$
\end{algorithmic}
\hrule
\begin{algorithmic}[1]
\fontsize{9.5}{14}\selectfont
\STATE \textbf{Initialize}
\STATE $k=0$
\WHILE{not leaving}
    \STATE $k = k+1$
    \STATE $\mathcal{D}_{[k, k+N]} =\mathrm{T^{3}P}(t_{[k-T,k]}, D_{[k-T,k-1]}, d_{[k-T,k]})$
    \STATE $\gamma_k = \text{AdapWght}(B_{[k-M,k]})$
    \STATE $\{R_k, \Delta t_k\} = \text{Srch}(\text{Opts}, \gamma_k, l, R_{k-1}, L_k, S_k, \mathcal{D}_{[k,k+N]})$
    \STATE Download chunk $k$ with bitrate $R_k$.
    \STATE Wait for $\Delta t_k$ seconds.
\ENDWHILE
\end{algorithmic}
\label{alg:alg1}
\end{algorithm}

\subsection{Overview}
Algorithm~\ref{alg:alg1} outlines the overarching structure of the BE-ABR system, which comprises two primary components.
We devise a \textit{Transformer-based, time-aware transmission delay predictor}, which we call $\mathrm{T^{3}P}$, that accurately forecasts the transmission duration of video chunks.
Leveraging the predictions from $\mathrm{T^{3}P}$, we propose a \textit{rate adaptation algorithm with fine-grained buffer control} to determine the download configuration (i.e., bitrate level and waiting time) of each chunk. 

\textbf{Transformer-based, time-aware transmission delay prediction.}
By leveraging historical bandwidth, sampling time, and video chunk information, $\mathrm{T^{3}P}$ predicts the download time for the next $N$ chunks at each bitrate level. 
Building upon the Transformer architecture, $\mathrm{T^{3}P}$ introduces a novel network structure that incorporates three specialized functional modules to effectively capture the evolutionary trends of irregular time-series network data. 
The details will be elaborated in Section~\ref{subsec:t3p}.

\textbf{Rate adaptation with fine-grained buffer control.}
Based on the current buffer status, historical data, and prediction results obtained from $\mathrm{T^{3}P}$, we collaboratively control the quality level and waiting time of each chunk to manage download bitrates and precisely regulate the buffer as well.
To achieve this, we first develop a \textit{network fluctuations based adaptive weighting} that dynamically adjusts the weight ratio between QoE and traffic wastage in the objective function (\ref{eq:16}) according to the bandwidth volatility, thereby enhancing the robustness of QoE performance.
Subsequently, we design a \textit{QoE-constraint optimal configuration searching} to quickly search for the optimal solutions of \textit{Problem 2} with guaranteed QoE loss rate.
Last, we implement the optimum strategy, downloading chunk $k$ with bitrate $R_k$ and waiting for $\Delta t_k$ seconds before repeating the process to determine configuration for the next chunk.
Further details are elaborated upon in Section~\ref{subsec:abr}.

\subsection{Transformer-Based, Time-Aware Transmission Delay Prediction}\label{subsec:t3p}
Forecasting the download time of transmitted chunks has always been a critical aspect of adaptive streaming systems, as it directly affects the calculation of rebuffering time, which is closely related to QoE performance.
In our context of wastage-controlled systems, the network predictor gains even more significance.
With decreased buffer occupancy, there is less room for error in throughput predictions, as there are not enough video segments stored in the buffer to compensate for inaccuracies. 
To address this need for highly precise transmission time predictions for each video chunk, we propose $\mathrm{T^{3}P}$ in this section.

\textcolor{myblue}{
\textbf{Challenges and solutions:}
In the following content, we first reveal the intricacies of the \textit{highly nonlinear complexity inherent} in the network information, as well as the \textit{irregular time series characteristics} present in historical network bandwidth sampling data. 
Then we introduce from a high-level perspective how the $\mathrm{T^{3}P}$ algorithm is designed to tackle these challenges.}

\textit{Highly non-linear complexity:} 
The relationship between available network information and transmission time exhibits high non-linearity and complexity. 
The user's allocated bandwidth is influenced by numerous cross-layer network factors\cite{fu2013survey} as well as the size of the requested chunk\cite{bartulovic2017biases,zhang2017modeling,yan2020learning,wang2023robust}. 
This complexity makes it difficult to effectively capture the features and patterns needed for accurate transmission time prediction.

To address this challenge, we first perform decoupling to separate the various influences on the allocated bandwidth, thereby simplifying the learning task.
Then we employ an encoder-only transformer to handle the pattern learning task.
As one of the most cutting-edge deep learning architectures built upon the attention mechanism, Transformers have demonstrated their efficacy in capturing complex patterns and dependencies \cite{ding2020hierarchical, zhou2021informer, devlin2018bert, antoun2020arabert, vaswani2017attention}, making them well-suited for handling non-linear and complex network data.

\begin{figure}[!t]
	\centering
	\includegraphics[width=3in]{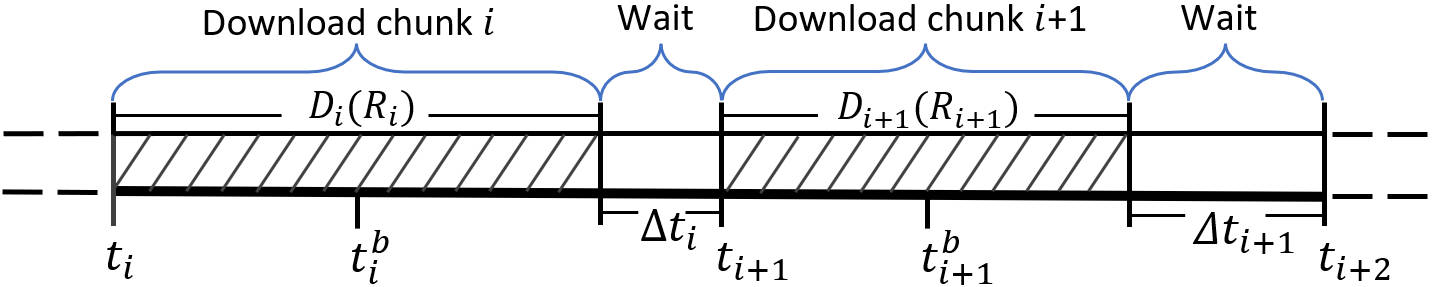}
	\caption{Process of video chunk downloading with key time points highlighted.}
	\label{fig2}
\end{figure}

\textit{Irregular time series:} The historical network data is essentially irregular time series data, as the measurements are not taken at fixed time intervals.
Figure~\ref{fig2} displays the download process and highlights the key time points, which helps to describe the irregular nature of network data sampling. 

By dividing the size of the $i$-th video chunk by its download time, we can obtain the average throughput $p_i$ during the download of the $i$-th chunk.
This value can be approximately considered as the bandwidth value at $t_i^{b}$, which is the midpoint of the download time for chunk $i$:
\begin{equation}
t_i^b  = t_i + \frac{1}{2} \times D_i(R_i).
\end{equation}
And the bandwidth sampling interval $\tau_i$, which refers to the time difference between consecutive observations of bandwidth, is calculated as:
\begin{equation}
\begin{aligned}
\tau_i &= t_{i+1}^b - t_{i}^{b} \\
&= t_{i+1} + \frac{1}{2} \times D_{i+1}(R_{i+1}) - t_i - \frac{1}{2} \times D_i(R_i) \\ 
&= \frac{1}{2} \times \left[\left.D_{i+1}(R_{i+1}) + D_i(R_i)\right.\right] + \Delta t_i.
\end{aligned}
\end{equation}
It is evident that the sampling intervals for the bandwidth are not constant.
This variability in time intervals could lead to changes in the correlation structures between data points.
Therefore, we incorporate sampling times as input features and design a time perception module based on a key-query attention mechanism \cite{luo2020hitanet} to effectively handle the irregular time-series network data.

\begin{figure}[!t]
	\centering
	\includegraphics[width=3.4in]{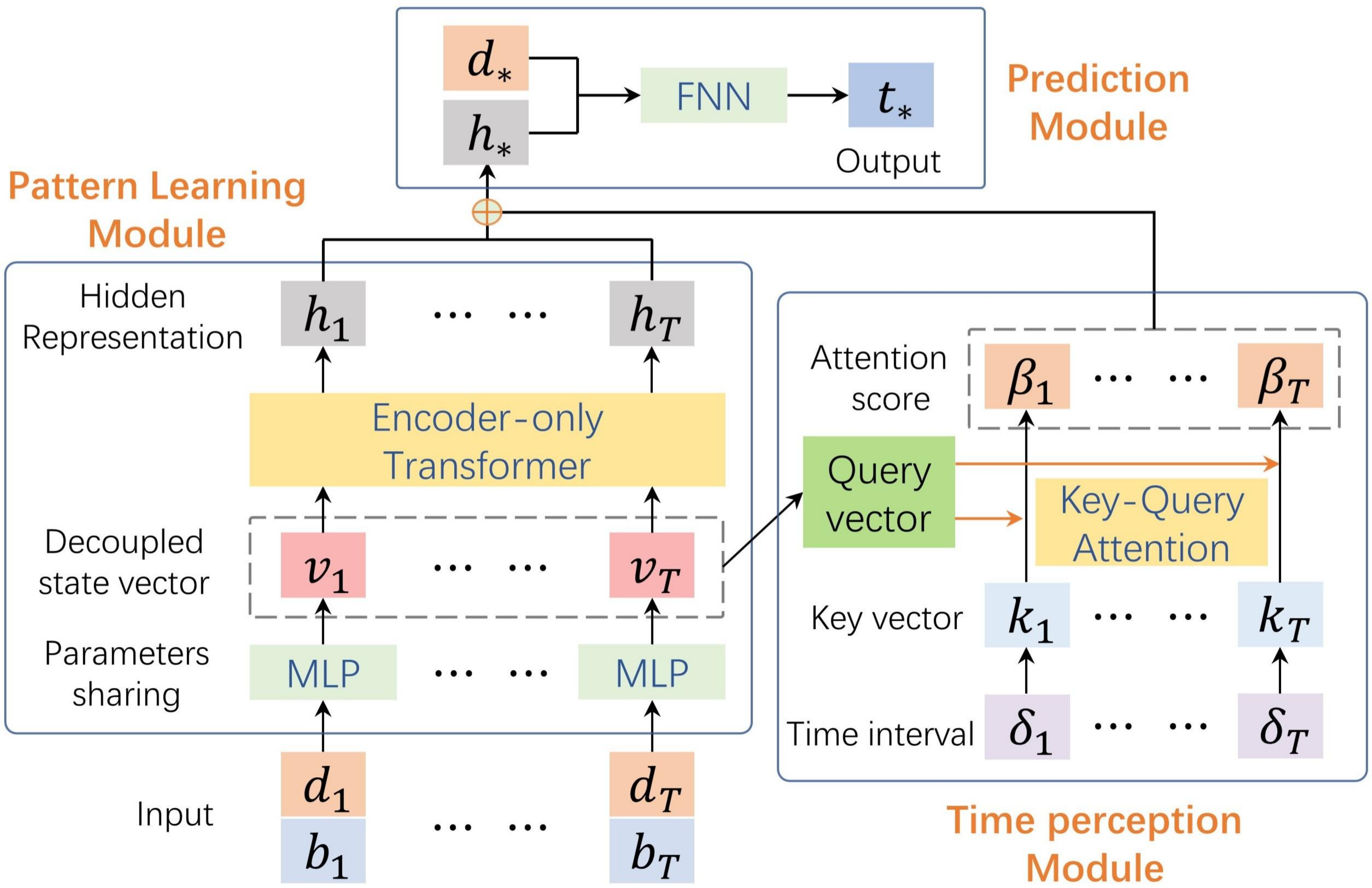}
	\caption{The structure of Transformer-based, time-aware transmission delay prediction model $\mathrm{(T^{3}P)}$.}
	\label{fig3}
\end{figure}

\begin{figure}[!t]
	\centering
	\includegraphics[width=2.7in]{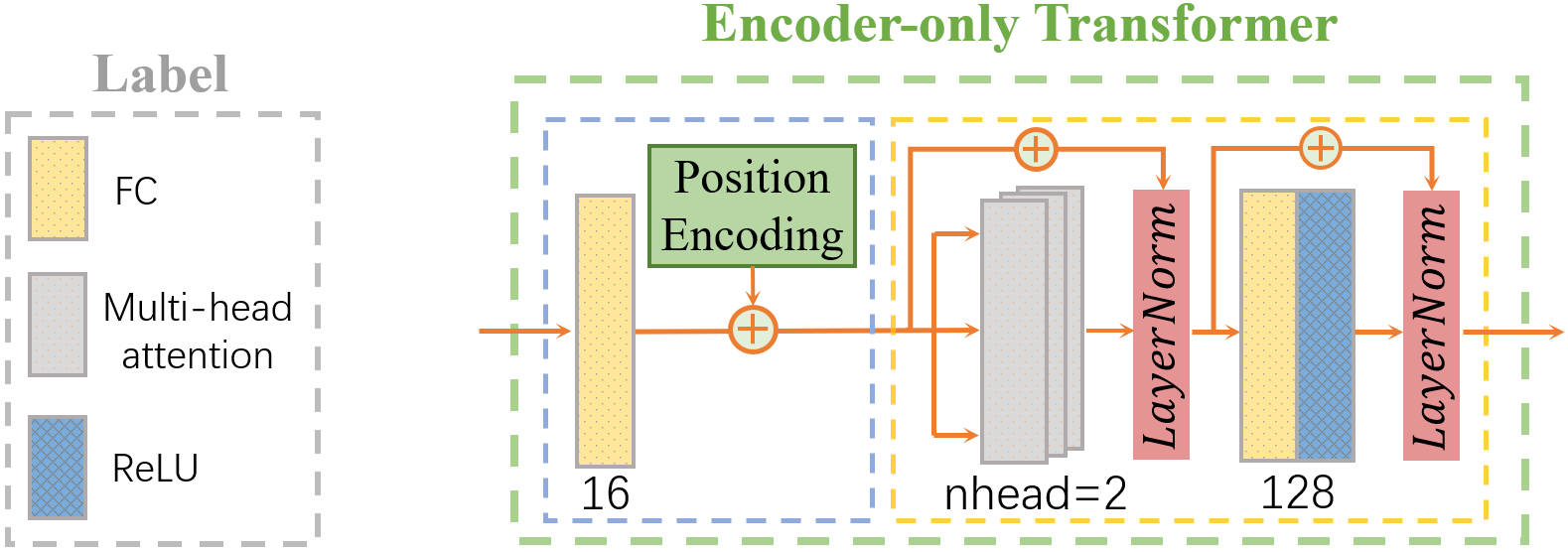}
	\caption{\textcolor{myblue}{The unfolding structure of the Encoder-only Transformer employed in $\mathrm{T^{3}P}$.}}
	\label{fig4}
\end{figure}

\textbf{Model:}
Considering the aforementioned network characteristics, we present $\mathrm{T^{3}P}$, a Transformer-based, time-aware transmission delay prediction model, to predict the transmission time of each chunk based on the historical measurements of network throughput, corresponding sampling time, and requested chunk size.

Figure~\ref{fig3} illustrates the basic network structure of $\mathrm{T^{3}P}$.
Our model consists of three main components: a pattern learning module, a time perception module and a prediction module.
Each module plays a crucial role in the overall efficacy of the model.
In the \textit{pattern learning module}, we first simplify the task to enable efficient network pattern capturing, which employs a Multilayer Perceptron (MLP) with shared parameters to decouple the influence of the requested chunk size.

Following this, we adopt an encoder-only transformer \cite{vaswani2017attention} architecture to learn the complex patterns of network performance and trends.
The primary objective of the \textit{time perception module} is to investigate the impact of the sampling time of historical data on network status prediction.
To achieve this, we employ a key-query attention mechanism that derives time-aware attention weights, evaluating the impact of varying sampling intervals on the temporal correlation between historical data at the same position in the sequence and the future bandwidth.
In the final step, the \textit{prediction module} integrates the outputs from both the pattern learning and time perception modules with the requested chunk size to provide a direct estimation of transmission time.

\textit{Pattern Learning Module}:
A variety of confounding factors determine the evolution of throughput, and even with powerful deep learning models, it is impossible to capture all influencing mechanisms simultaneously. 
Hence, we first simplify the learning task by decoupling the influence of requested chunk size on throughput.

We insert the historical $T$ data points $x_{k-T}, \cdots, x_{k-1}$ (we  use $x_1, x_2, \cdots, x_T$ instead for brevity) into the pattern learning module.
Specifically, $x_i$ is composed of $d_i$ and $b_i$, representing the size of chunk $i$ and the throughput measured at time $t_{i}^{b}$.
Owing to scheduling policies in network pipelines and TCP's slow start mechanism, allocated bandwidth is not linearly affected by chunk size.
Therefore, we respectively feed each $x_i$ into a MLP with shared parameters to obtain a decoupled network state vector $v_i \in IR^s$.
\textcolor{myblue}{
The MLP used here consists of three layers, each with 16 neurons. 
The first two layers use the Rectified Linear Unit (ReLU) activation function, while the last layer uses a linear one.}

Subsequently, we input $V = [v_1, v_2, \cdots, v_T]$ into an encoder-only Transformer model (denoted as $F$) to learn long-term dependencies among historical network states.
\textcolor{myblue}{
The detailed architecture of the Encoder-only Transformer model is illustrated in Fig.~\ref{fig4}.}
The model primarily consists of an input layer, a position encoding layer, and a standard encoder layer.
The input layer maps $V$ to a vector of dimension $d_{model}$ through a fully-connected network.
Then, the position encoding layer is added to incorporate temporal order information into the time series data, which is performed in the same manner as in Vaswani \textit{et al} .\cite{vaswani2017attention}.
Afterwards, we feed the resulting vector into a one-layer encoder to learn the long-term dependencies among the historical data.
The final output of our encoder-only Transformer is:
\begin{equation}
	[h_1, h_2, \cdots, h_T] = F([v_1, v_2, \cdots, v_T]).
\end{equation}

\textit{Time Perception Module:}
Due to the irregular temporal nature of network data, historical data at the same position in the sequence may exert varying impacts on the future bandwidth prediction. 
Therefore, it is essential to learn the influence of sampling time intervals on the correlation between data points.
In this module, we utilize a key-query attention mechanism \cite{luo2020hitanet} to generate time-aware attention weights.
By converting the decoupled network state vector as a query vector and the processed sampling time interval as a key vector, we can obtain attention weights that help adjust the significance of historical network data at different positions in the sequence with respect to the future bandwidth prediction.

First, we compute the interval $\delta_i$ between the historical bandwidth sampling time and the current time:
\begin{equation}
\delta_i = t_k - t_{i}^{b} 
\end{equation}
Then we convert each time interval $\delta_i$ into a key vector $k_i\in IR^s$:
\begin{equation}
	k_i = tanh(W_k \delta_i + b_k),
\end{equation}
where $W_k \in IR^s$ and $ b_k\in IR^s$ are parameters.
Meanwhile, we convert the historical network state vector $V$ into a query vector $q \in IR^s$:
\begin{equation}
	q = ReLU(V W_q + b_q),
\end{equation}
where $W_q \in IR^{T}$ and $b_q \in IR^s$ are parameters.

To investigate the influence of sampling time intervals on the correlation structures between data points, the key-query attention mechanism is applied to get an attention weight:
\begin{equation}
	\phi_i = \frac{q^T k_i}{\sqrt{s}}.
\end{equation}
We then get normalized time-aware attention weights $\beta$ through a softmax layer:
\begin{equation}
	\beta = Softmax(\phi) = [\beta_1, \beta_2, \cdots, \beta_T].
\end{equation}

Finally, we obtain the prediction of the hidden state vector of future network $h_*\in IR^s$ by combining the time-aware attention weights and the hidden representation of historical network states:
\begin{equation}
	h_* = \sum\limits_{i=1}^{T} \beta_i h_i.
\end{equation}

\textit{Prediction Module:}
This module is responsible for restoring the impact of requested chunk volume and generating the final prediction results for the transmission time.
First, we integrate $h_*$ with the size of the chunk being transmitted $d_*$.
Then, we employ a  multi-layer feedforward neural network (FNN) to restore the nonlinear effects of chunk size on allocated bandwidth.
\textcolor{myblue}{
The FNN consists of three layers with neuron distributions of 32-32-1, utilizing ReLU activation for the first two layers and linear activation for the final layer.}
The prediction result $t_*$ is expressed by:
\begin{equation}
	t_* = FNN(d_*, h_*).
\end{equation}

\textbf{Dataset and training:}
The performance of the $\mathrm{T^{3}P}$ relies heavily on the network data or network environment used for training.
Algorithms trained in simulated environments may perform poorly in the real world because it can still not emulate the diversity and dynamics of Internet paths well \cite{floyd2003internet,yan2018pantheon}.
To this end, we extract the desired training data from Puffer \cite{yan2020learning}.
Puffer is a free, publicly accessible video streaming service platform that streams six commercial television channels over the air.
Leveraging Puffer's network data logging, we construct a dataset comprising ten million data points. 
\textcolor{myblue}{This dataset is partitioned into a ratio of 8:1:1 for training, validation and testing respectively.}

We set the batch size as 512 and train our model on top of the PyTorch \cite{paszke2019pytorch} framework.
We adopt Mean Squared Loss as the loss function and use ADAM \cite{kingma2014adam} algorithm as the optimizer.
\textcolor{myblue}{To prevent overfitting, we employ early stopping with a patience of 7.}
We also use a warm-up learning rate schedule for smooth convergence, defined as:
\begin{equation}
	\begin{aligned}
		l_{rate} = & d_{model}^{0.5} \times min(step\_num ^{0.5}, \\ 
		& step\_num \times warmup\_steps ^{-1.5}).
	\end{aligned}
\end{equation}
Here, we set the $warmup\_steps$ to be 5000 empirically.

\subsection{Rate Adaptation with Fine-Grained Buffer Control}\label{subsec:abr}
Traditional ABR algorithms primarily control the buffer through bitrate selections, with the inter-chunk waiting time typically set to zero unless the buffer is full.
This means the download will not stop until the buffer reaches its full capacity.
However, bitrate selections alone provide limited control over the playback buffer.
This limitation arises as video servers typically provide a narrow range of bitrate options due to compression and storage expenses.
The situation worsens when the bandwidth greatly exceeds the highest bitrate over an extended duration.

Fine-grained buffer control is crucial for attaining high-quality and cost-efficient streaming experiences.
It allows precise control of the buffer to reach an optimal level that balances QoE performance and bandwidth conservation.
Therefore, in this study, we propose treating both the waiting time ${\Delta t_k}$ and bitrate ${R_k}$ as optimization variables, controlling them cooperatively to achieve precise buffer control.

\textbf{Network fluctuations based adaptive weighting:} In adaptive streaming systems, QoE performance is heavily reliant on the precision of transmission time predictions, as this is essential for preventing rebuffering events.
Intuitively, providing accurate estimates in environments with substantial throughput variability is much more challenging.
These variations may result from a range of factors, including network congestion, changes in network infrastructure, and interference from other devices or users sharing the same network resources, which are difficult or even impossible to predict.

In order to maintain stable and robust high QoE performance, we dynamically adapt the weight ratio between QoE and traffic wastage in the objective function (\ref{eq:16}) based on real-time bandwidth volatility at each time step.
For instance, when bandwidth fluctuations are severe, the system is more susceptible to rebuffering due to imprecise predictions, which could lead to a poor video viewing experience.
Consequently, under such conditions, we assign a greater weight to the QoE factor.

In particular, we use $CV_k$, the coefficient of variation of the past $M=5$ throughput, to reflect current bandwidth fluctuations.
The coefficient of variation (CV) \cite{CV} is a widely used statistical measure that quantifies the extent of variability in data.
\textcolor{myblue}{Calculated as the standard deviation divided by the mean value of the sample, the CV coefficient increases with the variability of the data, and its range of values is $(0, +\infty]$.}
Denoting the average bandwidth during the downloading process of the last $M$ chunks as $\overline{B}_{[k-M,k]}$, the formula for calculating $CV_k$ is given by:
\begin{equation}
CV_{k} = \frac{\sqrt{\displaystyle\sum_{i=k-M}^{k-1}(b_{i}-{\overline{B}_{[k-M,k]}})^{2}\big/(M-1)}}{\displaystyle\sum_{i=k-M}^{k-1}b_i / M}.
\end{equation}

\textcolor{myblue}{Then, we apply an exponential mapping to $CV_k$, and take its reciprocal, to generate the adjustment factor $\gamma_k$.}
The specific calculation formula is:
\begin{equation}
    \gamma_k = \frac{1}{\text{exp}({CV}_k)}.
\end{equation}
\textcolor{myblue}{
The value of $\gamma_k$ lies within the range $(0,1]$.
When the bandwidth fluctuation is significant, the value of $\gamma_k$ tends to 0. When the bandwidth is stable, the value of $\gamma_k$ tends to 1. 
By applying $\gamma_k$ enables us to adjust the proportion of the waste term in the optimization target based on network fluctuations.}
At time $t_k$, the reward function that we aim to maximize is re-expressed as:
\begin{equation}
	\begin{aligned}
		AdapRew_k^{k+N} &= QoE_k^{k+N} - \gamma_k \times \beta \times WAS_k^{k+N}.
	\end{aligned}	
\end{equation}

\begin{algorithm}[!t]
\caption{QoE-Constraint Optimal Configuration Searching.}
\begin{algorithmic}
\renewcommand{\algorithmicrequire}{ \textbf{Comments:}}
\fontsize{9.5}{14}\selectfont
\REQUIRE
\STATE GA: Apply Genetic Algorithm to searching for the
optimal solutions.
\STATE FindMaxQoE: Compute the expected $QoE_{k}^{k+N}$ of all the options and return the maximum value.
\end{algorithmic}
\hrule
\begin{algorithmic}[1]
\fontsize{9.5}{12}\selectfont
\REQUIRE $\text{Opts}, \gamma_k, l, R_{k-1}, L_k, S_k, \mathcal{D}_{[k,k+N]}$  \\
\ENSURE $R_k, \Delta t_k$
\STATE $maxQoE = \text{LookupTable}(\text{Opts}, R_{k-1}, L_k,S_k, \mathcal{D}_{[k,k+N]})$
\STATE $bound = l \times maxQoE$
\STATE $\{R_k,\cdots, R_{k+N}, \Delta t_k, \cdots, \Delta t_{k+N}\} = \text{GA}(\text{Opts}, bound, $ \\
\hfill $ \gamma_k, R_{k-1}, L_k, S_k, \mathcal{D}_{[k,k+N]})$
\STATE \textbf{return} $R_k, \Delta t_k$
\end{algorithmic}
\label{alg:alg2}
\end{algorithm}

\textbf{QoE-constraint optimal configuration searching:}\label{subsec:QcSS}
Although QoE and bandwidth savings are not entirely conflicting objectives, reducing traffic wastage may inadvertently lead to QoE degradation in certain instances.
Nevertheless, QoE performance is of such paramount importance that most video providers can tolerate minimal or no decline. 
Consequently, we develop a protection mechanism for the QoE loss ratio $l$, \textcolor{myblue}{which is defined as the proportion of QoE loss incurred due to efforts to reduce data wastage.}
This ratio can be adjusted between 0 and 1 to cater to the specific requirements of video providers.

As shown in Algorithm~\ref{alg:alg2}, we first compute the maximum expected $QoE_{k}^{k+N}$ achievable at time step $t_k$.
\textcolor{myblue}{
By setting the coefficient $\beta = 0$ for the $WAS_{k}^{k+N}$ term in the optimization objective of \textit{Problem 2}, the problem reverts to the fundamental task of maximizing QoE.
The solution to this optimization problem provides the maximum achievable QoE scores without considering wastage.
Here, we adopt a table enumeration approach to solve this optimization problem.
Specifically, We first enumerate and solve each state offline.
During the online phase, we directly read the optimal solution from the pre-calculated table, which has almost no computation delay. 
We employ the same table compression techniques as in fastMPC~\cite{yin2015control}.
The lower limit of acceptable QoE scores is then derived by multiplying the maximum expected $QoE_{k}^{k+N}$ with the predetermined QoE loss ratio $l$.} 
Strategies that have expected QoE scores below this threshold will not be adopted.
With this constraint, which limits the loss ratio of QoE performance, we proceed to solve the maximization problem in \textit{Problem 2} with Genetic Algorithm (GA).

\textcolor{myblue}{
In \textit{Problem 2}, the state space is defined by the following dimensions: (1) the length of the currently buffered video, (2) the volume of data currently buffered, (3) the previously chosen bitrate, and (4) the predicted transmission time for the next $N$ chunks.
The complexity of this high-dimensional state space makes traditional look-up table methods impractical.
The Genetic Algorithm (GA) \cite{holland1992adaptation}, inspired by principles of natural selection and genetic evolution, is a potent optimization search technique. 
The parallelism of GA allows it to handle multiple possible solutions in the solution space simultaneously, greatly enhancing the search efficiency.
Its proficiency in discovering optimal solutions within intricate problem spaces makes it a fitting choice for our problem.}
Empirically, we set the parameters for GA as follows: $size\textunderscore pop=50, max\textunderscore iter=200 ,prob\textunderscore mut=0.001, precision=0.1, early\textunderscore stop=5$.

By employing the Genetic Algorithm, we rapidly search and obtain the near-optimal configuration for the next $N$ chunks: $\{R_k, \dots, R_{k+N}, \Delta t_k, \dots, \Delta t_{k+N}\}$.
This download strategy achieves minimization of traffic waste while maintaining the actual QoE loss below the predefined ratio $l$.
Lastly, we provide $\{R_k, \Delta t_k\}$ to the controller, and proceed to download chunk $k$ with bitrate $R_k$ and waiting for $\Delta t_k$ seconds.

\section{Experimental results}
First, we compare BE-ABR against several baselines with real prototype implementations on commodity WiFi and 4G/LTE networks in the wild.
Our large-scale evaluations demonstrate the effectiveness of BE-ABR in improving QoE performance and reducing traffic wastage.
Secondly, we undertake a component-wise study to analyze the contributions of and impacts on the overall performance of the two critical designs proposed in BE-ABR.
\textcolor{myblue}{
Further, a user study is conducted to evaluate the performance of BE-ABR with actual user engagement.
We also assess the generalization ability of BE-ABR across multiple real-world traces.}
Lastly, we examine the predictive capability of our $\mathrm{T^{3}P}$ model using multiple evaluation metrics on a massive dataset of real-world network data.

\subsection{Experimental Setup}
We describe the experiment settings in our evaluation.

\textbf{Testbed:} \label{subsec:exp_setup}
Given that available bandwidth is constantly influenced by users' real-time downloading activities, experiments conducted over fixed throughput traces cannot fully reflect these network conditions.
Therefore, our primary choice is to carry out experiments in real-world network environments.
We establish a real video streaming testbed to compare the performance of BE-ABR with that of previous approaches in the wild.

\textcolor{myblue}{In our experimental setup, we use FFmpeg~\cite{ffmpeg} to pre-encode test videos in High Efficiency Video Coding (HEVC) format, and we apply MP4Box~\cite{MP4box} to package and dashify HEVC bitstreams.}
The test videos are stored on an HTTP web server, which is implemented on an Ubuntu 20.04 system.
The video client interfacing with the server is a Dell OptiPlex 3080MFF Mini PC.
It comes equipped with an Intel UHD Graphics 610 mobile-class GPU, an Intel Pentium G6405T 1.8 GHz dual-core processor, 4 GB of RAM, and a 128 GB SSD, similar to the computing resources of a typical smartphone.
\textcolor{myblue}{Following Pensieve~\cite{mao2017neural}, we run $\mathrm{T^{3}P}$ on a standalone ABR server, with a communication latency of around 20 ms to the mobile device. 
However, if the mobile device has sufficient computational capabilities, it is also feasible to deploy the computation of $\mathrm{T^{3}P}$ on the device itself.
We modify dash.js~\cite{dash} to support BE-ABR as well as comparison algorithms.}

The real-world experiment is conducted in two distinct network environments: a China Telecom’s 4G/LTE cellular network at a coffee shop and a public Wi-Fi network at a school.
\textcolor{myblue}{Table~\ref{tab:network_statics} presents the key statistics of the network environments.}

\begin{table}[!tbp]
    \footnotesize
    \centering
    \caption{\textcolor{myblue}{Statistics of real-world networks}}
    \label{tab:network_statics}
    \arrayrulecolor{myblue}
    \begin{tabular}{cccc}
        \toprule
        & \textcolor{myblue}{\makecell{Average \\ Throughput}} & \textcolor{myblue}{\makecell{Coefficient of \\ Variation}} & \textcolor{myblue}{\makecell{Range of \\ Throughput}} \\
        \midrule
        \textcolor{myblue}{Wi-Fi} & \textcolor{myblue}{3.55 Mbps} & \textcolor{myblue}{0.32} & \textcolor{myblue}{0.1-6.8} \\
        \textcolor{myblue}{4G/LTE} & \textcolor{myblue}{3.94 Mbps} & \textcolor{myblue}{0.56} & \textcolor{myblue}{0.2-10.3} \\
        \bottomrule
    \end{tabular}
\end{table}

\textbf{Video test sets:}
\textcolor{myblue}{
We utilize two distinct video sets with varied bitrate ladders for testing.
1) \textit{SD Video Set}:} This collection comprises fifteen standard definition videos.
Each video is encoded at the following bitrate levels: $\mathcal{R} = \{350, 600, 1000, 2000, 3000\}$ Kbps, which corresponds to common video modes in $\{240, 360, 480, 720, 1080\}$ p.
\textcolor{myblue}{
2) \textit{UHD Video Set}: This series contains five ultra-high-definition videos.
They are encoded as \{0.2, 0.4, 0.8, 1.2, 2.2, 3.3, 5.0, 6.5, 8.6, 10, 12, 16, 20\} Mbps, which replicates the bitrates sets in\cite{zhang2021post}.}
All the test videos are encoded using the HEVC codec.
The average length of the test videos is 350 seconds, with chunk duration of 2.133s and 2.667s.

\textbf{Algorithms for comparison:}
We do contrast experiments with the following 5 baselines which collectively represent the exemplary adaptive streaming techniques:
\begin{itemize}
    \item \textit{MPC:}
    MPC \cite{yin2015control} plans a trajectory of bitrates over a limited horizon of 5 chunks to maximize their expected QoE based on the buffer occupancy and throughput prediction.
    \item \textit{RobustMPC:}
    RobustMPC enhances the robustness upon MPC by taking prediction error into account and optimizing the worst-case QoE. The prediction error is computed as the max absolute percentage error of the past 5 chunks.
    \item \textit{BBA:}
    Proposed by Huang \textit{et al.} \cite{huang2014buffer}, BBA provides a buffer-based approach for rate adaptation.
    It selects bitrates with the purpose of maintaining the buffer occupancy within a suitable range.
    In our experiment, we adopt the function suggested by Huang and set the size of the reservoir and cushion to 5s and 10s, respectively.
    \textcolor{myblue}{
    \item \textit{BOLA}:
    Another buffer-based ABR scheme proposed by Spiteri \textit{et al.}\cite{spiteri2020bola}.
    BOLA frames the bitrate adaptation problem as a utility maximization problem and resolves it using the Lyapunov function.
    We set the key parameters for BOLA as suggested in \cite{spiteri2020bola}.}
    \item \textit{Fugu:}
    Fugu \cite{yan2020learning} trains a DNN-based transmission time predictor with data collected from a real publicly accessible website, and informs a MPC-based controller to make bitrates decisions for the upcoming chunks.
    \textcolor{myblue}{
    \item \textit{Pensieve:}
    Mao \textit{et al.} \cite{mao2017neural} propose a reinforcement learning ABR scheme, Pensieve, which trains a neural network to select bitrates for future video chunks.
    We use the pre-trained model.}
    \item \textit{PSWA:} PSWA \cite{zhang2021post} presents a framework to reduce data wastage by regulating the aggressiveness of bitrate selection and the buffer limit of the existing ABR algorithm.
    We consider the RobustMPC under the PSWA framework as the representative method of PSWA for comparison.
\end{itemize} 

\begin{figure}[!tbp]
    \centerline{\includegraphics[width=2in]{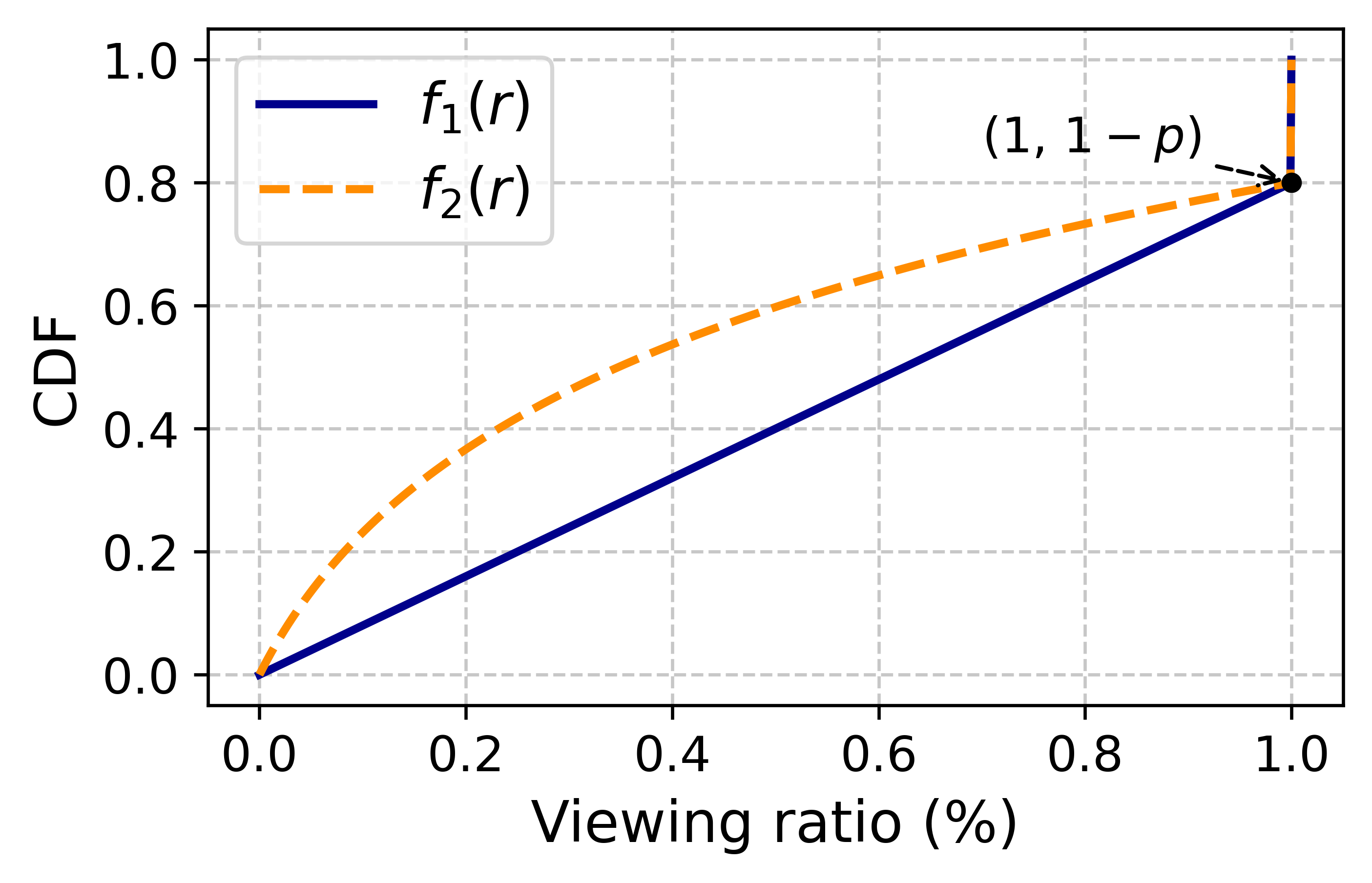}}
    \caption{CDF of viewing ratio under two assumed user departure behavior modes.}
    \label{pattern}
\end{figure}
\textbf{Departure behavior simulations:}
To the best of our knowledge, there is no public real-world datasets recording the user's viewing behavior during streaming.
\textcolor{myblue}{
Hence, we create two hypothetical scenarios to evaluate the extent of traffic wastage.}

In the first scenario, we assume that the user may depart at any moment, with the probability being $1-p$.
Here, $p$ is the probability that the user watched the video from beginning to end.
By denoting the ratio of the viewed video length to the full video length by $r$, 
the cumulative distribution function $f_1(r)$ for this scenario is as follows:
\begin{equation}
	f_1(r)=
\begin{cases}
	(1-p) * r, &r \in [0,1) \vspace{2ex}\\
	1,  &r = 1
\end{cases}
\end{equation}

The second scenario assumes that users are more likely to leave at the beginning of the streaming session. 
This presumption aligns with typical viewer behavior: users typically preview a new video before deciding whether to continue watching.
This assumption's validity is further supported by the statistical analysis of real user data in \cite{li2023dashlet}.

As such, we employed a logarithmic function to construct $f_2(r)$:
\begin{equation}
	f_2(r)=
	\begin{cases}
		(1 - p) * \log_{1+a}(1 + ar), &r \in [0,1) \vspace{2ex}\\
		1,  &r = 1
	\end{cases}
\end{equation}
Here, the coefficient $a$ modulates the probability of a user leaving during the browsing phase.
The larger the value of $a$, the higher the probability that a user will leave at the browsing phase.
In this function, the probability that the user will not leave until the video ends is still $p$.
The Cumulative Distribution Function (CDF) curve of viewing ratio under two assumed user departure behaviors are illustrated in Fig.~\ref{pattern}.

In our experiments, we set the parameters to be $p=0.2$ and $a=10$.
\textcolor{myblue}{
Before each streaming test, we randomly generate a user departure time based on the above distribution. 
During the streaming test, we halt transmission when this time point is reached, simulating a user's exit.
We then record the size of remaining video data in the buffer and calculate the QoE scores, which serve as the performance metric for this streaming session.}

\textbf{QoE metrics:}
\textcolor{myblue}{
In this paper, we employ two kinds of QoE metrics to evaluate user experience quality.
The first is $QoE_{lin}$, which is calculated in the same manner as\cite{yin2015control, mao2017neural, yan2020learning}.}
The function is defined as follows:
\begin{equation}
\begin{aligned}
    QoE_{lin} &= \alpha_1\times \sum_{i=1}^{N_1}R_i - \alpha_2 \times \sum_{i=1}^{N_2} Reb_i \\
    &- \alpha_3 \times \sum_{i=2}^{N_1}|R_{i} - R_{i-1}|
\end{aligned}
\end{equation}
\textcolor{myblue}{
In this equation, $Reb_i$ is the buffering duration during chunk $i$ download, $N_1$ represents the number of fully viewed chunks, and $N_2$ stands for the number of chunks already downloaded.
Here, we take into calculation only the chunks that have actually been viewed by the user, as these are the ones that genuinely affect user experience, rather than all downloaded chunks.} 
In $QoE_{lin}$, the perceived quality is quantified using the bitrate of the chunks.
And the weights are set as follows: $\alpha_1=\alpha_3=1, \alpha_2=600$.

\textcolor{myblue}{
The second metric used is $QoE_{log}$, as employed in~\cite{mao2017neural, spiteri2020bola}.
The function is defined as:}
\begin{equation}
\color{myblue}{
\begin{aligned}
    QoE_{log} &= \alpha_1\times \sum_{i=1}^{N_1}ln(R_i/R_{min}) - \alpha_2 \times \sum_{i=1}^{N_2} Reb_i \\
    &- \alpha_3 \times \sum_{i=2}^{N_1}|ln(R_{i}) - ln(R_{i-1})|
\end{aligned}}
\end{equation}
\textcolor{myblue}{
$QoE_{log}$ uses $ln(R_i/R_{min})$ to measure the quality of the video chunks. 
This simulates a scenario where the marginal improvement in perceived video quality tends to diminish as the bitrate increases.
The weights are set the same as in~\cite{mao2017neural}: $\alpha_1=\alpha_3=1, \alpha_2=266$.}

\begin{figure*}[!t]
    \centering
    \captionsetup[subfloat]{font=footnotesize}
    \subfloat[WiFi \& $f_1$ mode]{\includegraphics[width=0.24\textwidth]{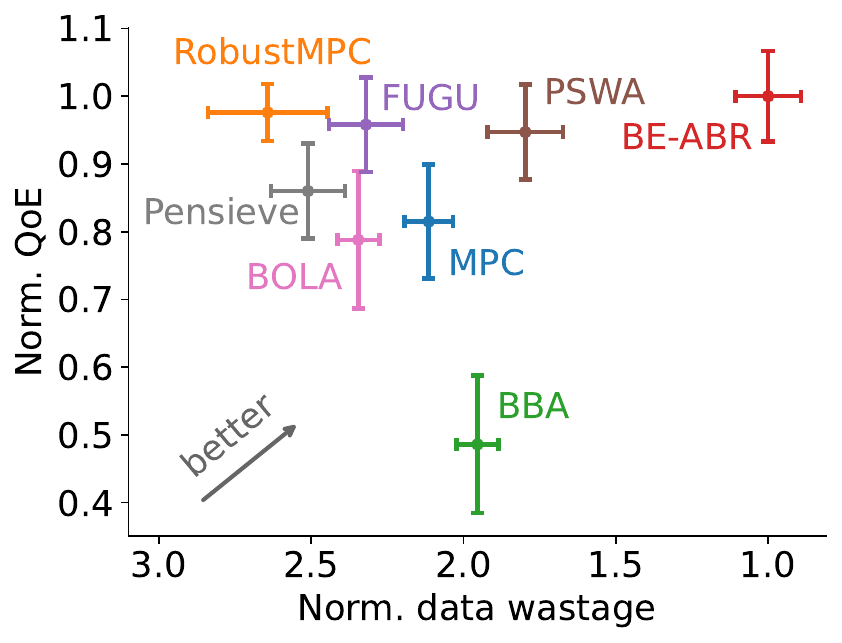}%
        \label{fig:fig3a}}
    \hfill
    \subfloat[WiFi \& $f_2$ mode]{\includegraphics[width=0.24\textwidth]{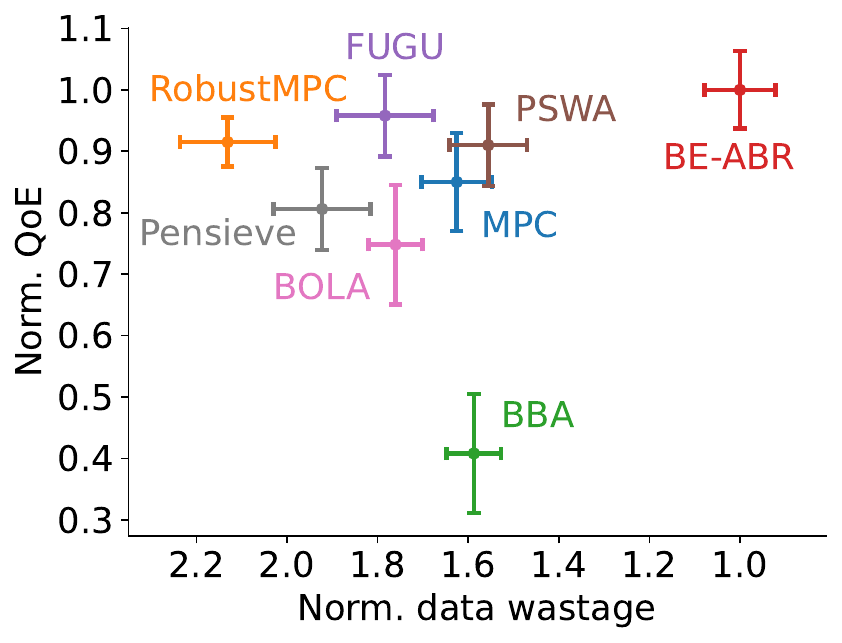}%
        \label{fig:fig3b}}
    \hfill
    \subfloat[4G/LTE \& $f_1$ mode]{\includegraphics[width=0.24\textwidth]{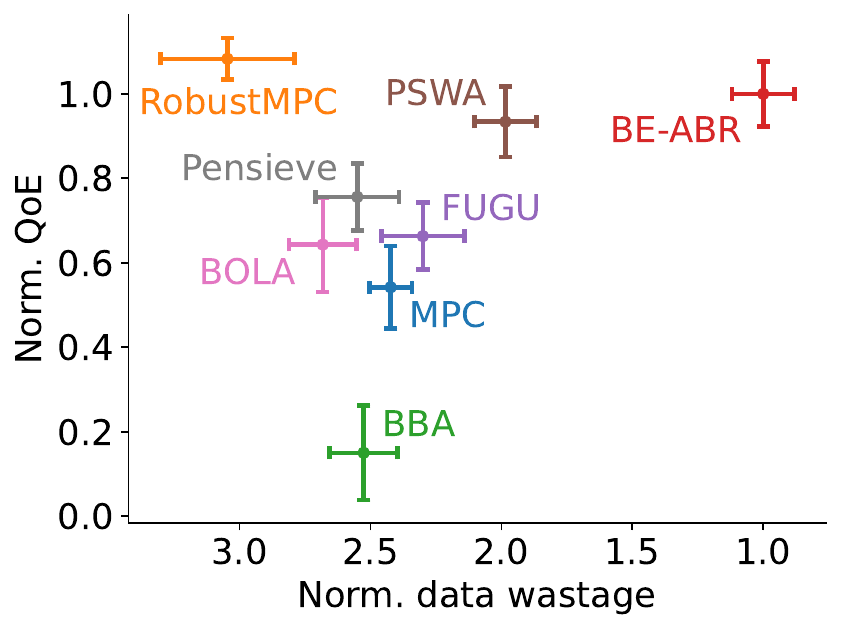}%
        \label{fig:fig3c}}
    \hfill
    \subfloat[4G/LTE \& $f_2$ mode]{\includegraphics[width=0.24\textwidth]{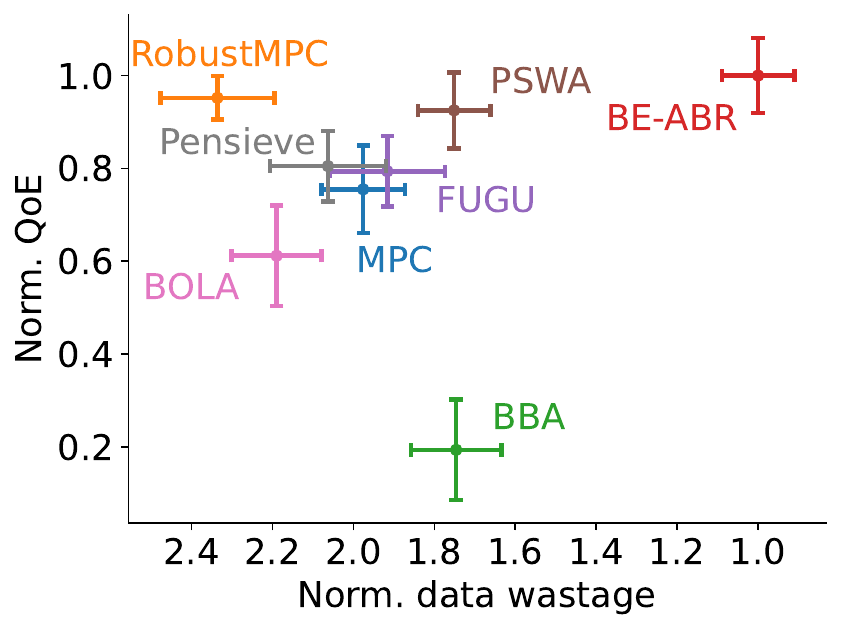}%
        \label{fig:fig3d}}
    \caption{Comparison of traffic wastage and QoE performance under two user behavior modes over commodity WiFi and 4G/LTE networks in the wild. Metrics are normalized against the performance of BE-ABR. \textcolor{myblue}{QoE metrics are computed as $QoE_{lin}$.}}
    \label{fig:fig3}
\end{figure*}

\begin{figure*}[!t]
	\centering
        \captionsetup[subfloat]{font=footnotesize}
	\subfloat[WiFi \& Quality]{\includegraphics[height=0.7in]{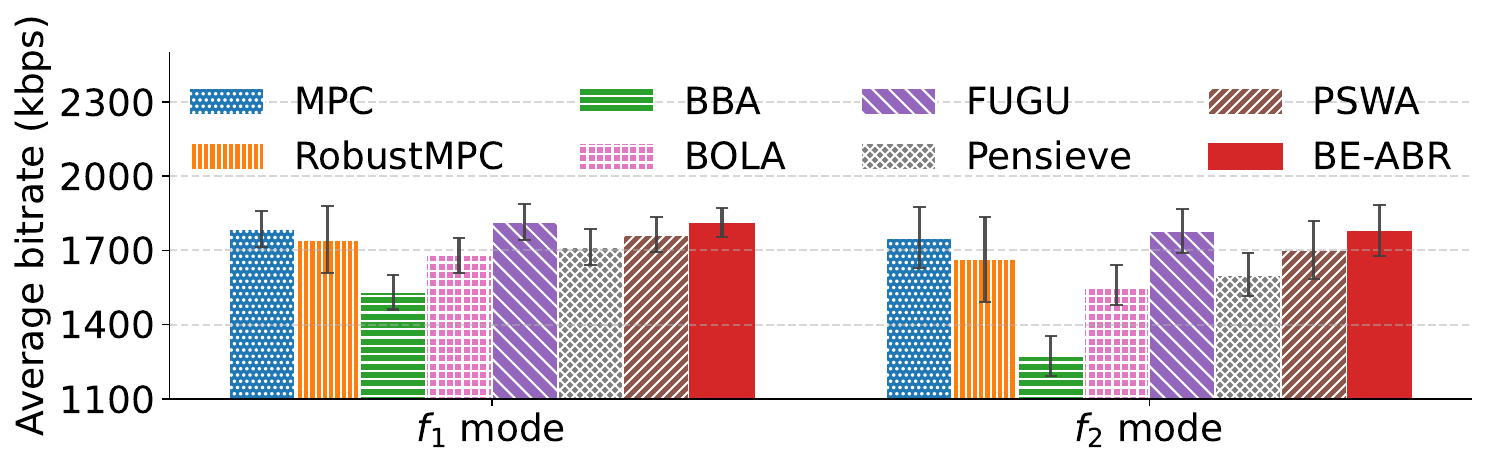}%
		\label{fig:fig4a}}
	\hfil
	\subfloat[WiFi \& Rebuffer ratio]{\includegraphics[height=0.7in]{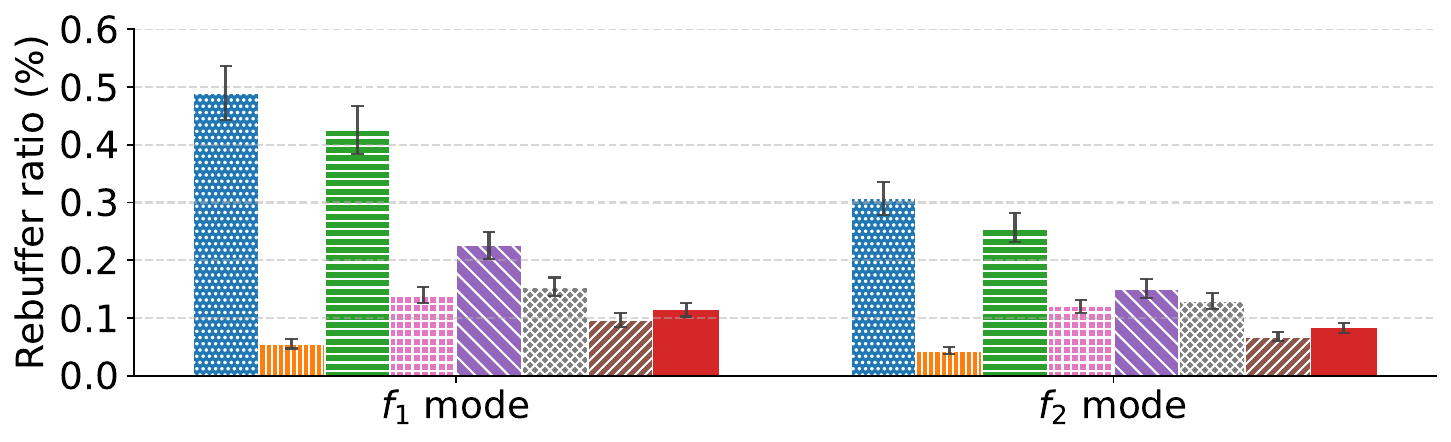}%
		\label{fig:fig4b}}
	\hfil
	\subfloat[WiFi \& Switch]{\includegraphics[height=0.7in]{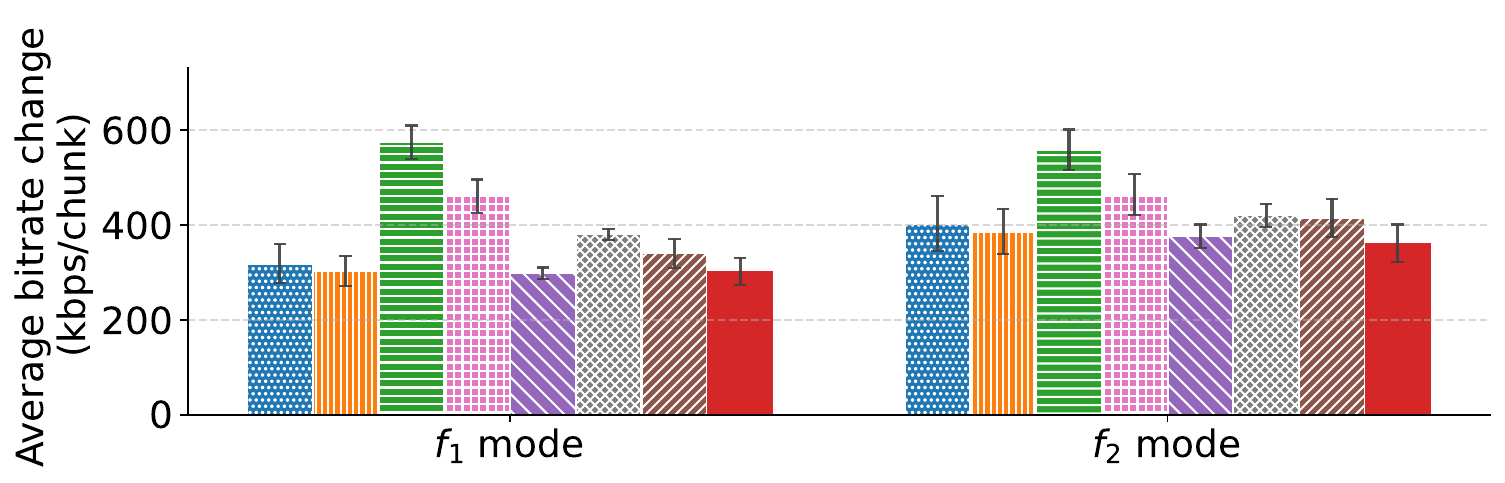}%
		\label{fig:fig4c}}
	\hfil
	\subfloat[4G/LTE \& Quality]{\includegraphics[height=0.7in]{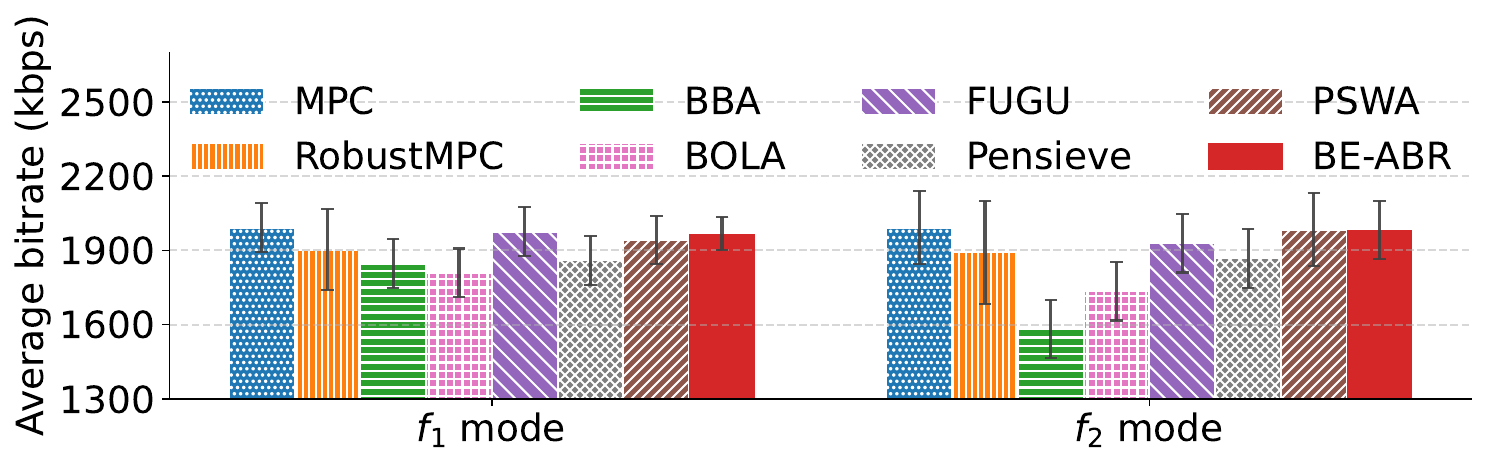}%
		\label{fig:fig4d}}
	\hfil
	\subfloat[4G/LTE \& Rebuffer ratio]{\includegraphics[height=0.7in]{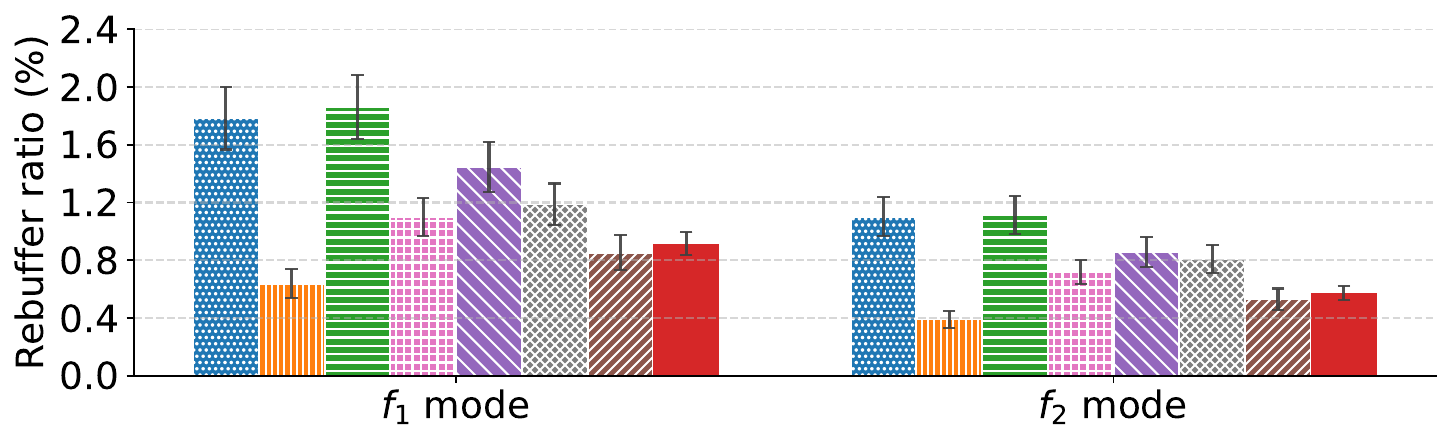}%
		\label{fig:fig4e}}
	\hfil
	\subfloat[4G/LTE \& Switch]{\includegraphics[height=0.7in]{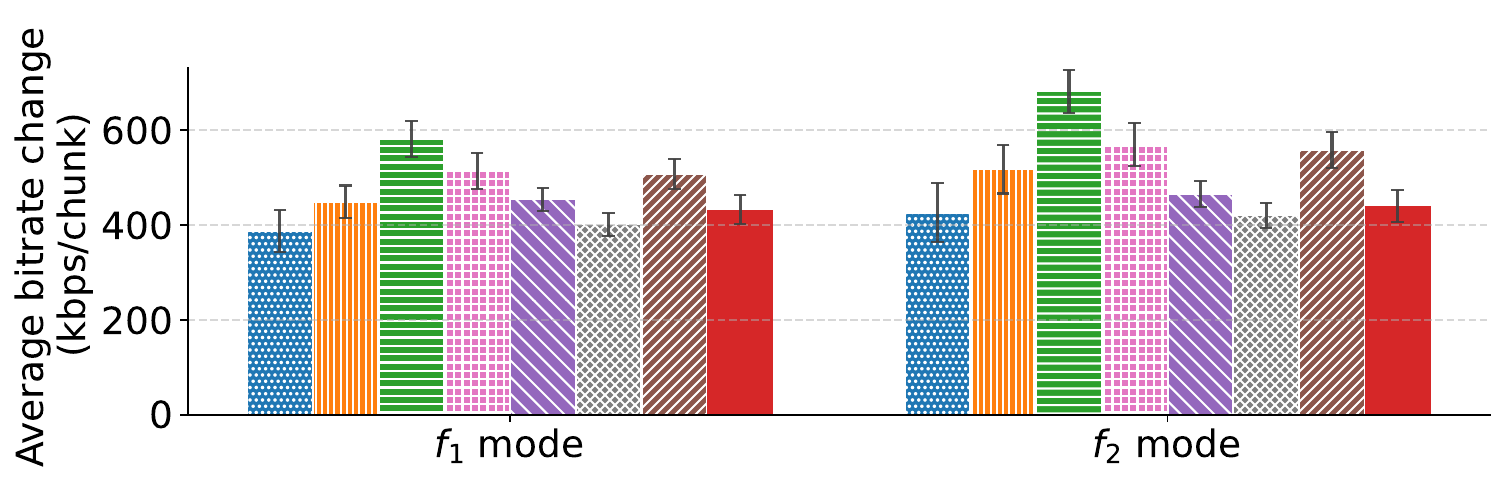}%
		\label{fig:fig4f}}
	\hfil
	\caption{Comparison of QoE-related assessment indexes under two user behavior modes over commodity WiFi and 4G/LTE networks in the wild.}
	\label{fig:fig4}
\end{figure*}

\begin{figure}[!t]
	\centering
        \captionsetup[subfloat]{font=footnotesize}
	\subfloat[case \uppercase\expandafter{\romannumeral1}]{\includegraphics[width=1.72in]{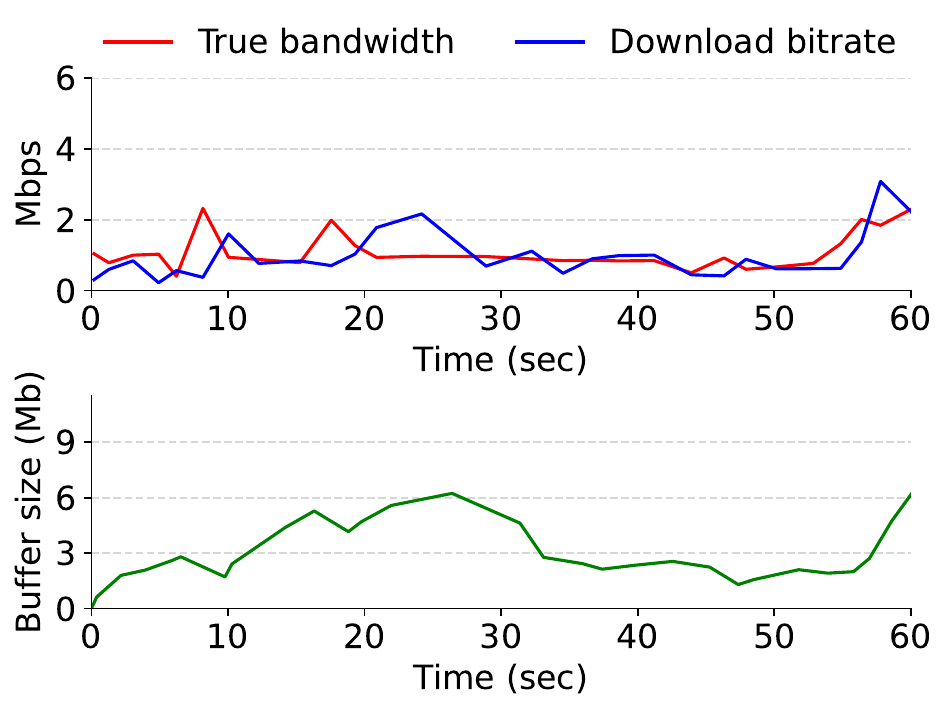}%
		\label{fig:fig70a}}
	\hfil
	\subfloat[case \uppercase\expandafter{\romannumeral2}]{\includegraphics[width=1.72in]{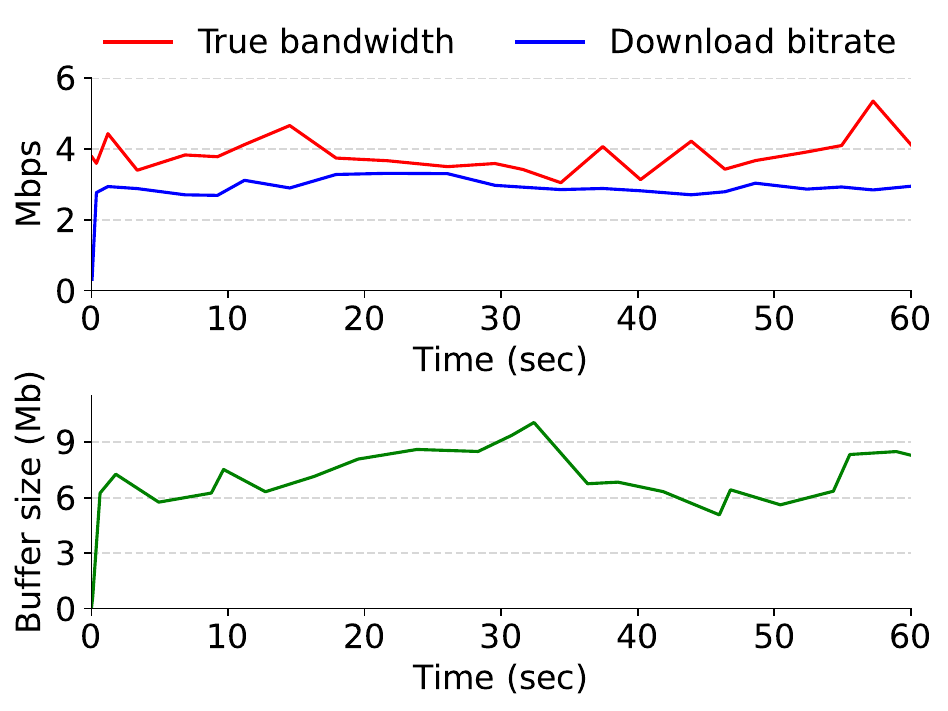}%
		\label{fig:fig70b}}
	\caption{\textcolor{blue2}{A case study showcasing bitrate selections and buffer dynamics of BE-ABR. Both cases were extracted from WiFi environment test results.}}
	\label{fig:fig70}
\end{figure}

\subsection{Improvement over Real Wireless Network Links}
We conduct real-world experiments using the \textit{SD Video Set}.
Within each network, we run the test videos twenty time for each ABR algorithm.
Figure~\ref{fig:fig3} illustrates the overall performance of each scheme under diverse user behaviors on real-world WiFi and 4G/LTE networks.
In the case of WiFi, BE-ABR surpasses several baselines in both QoE and traffic wastage metrics.
Specifically, BE-ABR attains the highest QoE scores and reduces traffic wastage by 35.7\%$\sim$62.2\%.
Regarding 4G/LTE results, BE-ABR outperforms the other five algorithms and closely matches RobustMPC in QoE performance.
Moreover, it continues to yield minimal traffic wastage, which is reduced by 37.5\%$\sim$67.2\%.

In conjunction with Fig.~\ref{fig:fig4}, which displays the detailed values of three QoE indicators, we provide a more in-depth analysis of the aforementioned results.
Firstly, regardless of whether on a highly volatile LTE network or a relatively stable WiFi network, BE-ABR significantly reduces the amount of wasted data.
\textcolor{myblue}{
Compared to the two buffer-based approaches, i.e., BBA and BOLA, BE-ABR successfully reduced data wastage by 36.9\%-62.7\%.
Other hybrid schemes also generate significantly more traffic wastage than BE-ABR, with FUGU at 1.78-2.32$\times$, Pensieve at 1.92-2.55$\times$, and MPC at 1.63-2.42$\times$.}
During non-peak hours, they continuously download video chunks until reaching the upper limit.
In this scenario, the buffer occupancy maintains a relatively stable state and fluctuates slightly below the limit.
When users leave during this period, substantial video data remaining in the buffer goes to waste.

RobustMPC employs a conservative forecast of future bandwidth, causing the actual download time to be generally shorter than expected.
As a consequence, the average buffer length of RobustMPC is greater than that of the other algorithms, making it come top in traffic wastage.
PSWA selects wastage-aware parameters for each bandwidth level through an offline analysis of historical transmission data, thereby reducing traffic wastage to a certain degree.
However, due to the absence of buffer perception and accurate predictions of future bandwidth, the selected parameters may not be optimal, causing a notable disparity with BE-ABR in terms of bandwidth savings.

Second, BE-ABR achieves the best QoE performance under WiFi networks and enhances QoE scores by up to 2.45$\times$.
This improvement can be primarily attributed to BE-ABR's low rebuffer ratio, as shown in Fig.~\ref{fig:fig4b}, ranging from 0.083$\%$ to 0.114$\%$.
The WiFi environment is generally stable, exhibiting fewer fluctuations and a reduced likelihood of sudden network changes.
With less uncertainty and fewer unpredictable factors in WiFi conditions, our Transformer-based $\mathrm{T^{3}P}$ effectively predicts the network's evolution trend.
\textcolor{myblue}{
In comparison to Pensieve and FUGU, which also employ learning-based methods, BE-ABR boosts QoE performance by a margin of 16.3\%-50.8\%.}
RobustMPC also achieves a low rebuffer ratio by adopting a conservative throughput estimation.
However, this comes at the expense of sacrificing some video quality, which is a crucial factor in video services.

The third finding is, under 4G networks, BE-ABR exceeds the other six algorithms on QoE scores but slightly underperforms compared to RobustMPC.
Predicting bandwidth for wireless networks is much harder than for WiFi networks, since the wireless channel features highly random characteristics.
User activity, the shared nature of the wireless medium, and signal blockages all contribute to the high volatility of mobile wireless connections.

As depicted in Fig.~\ref{fig:fig4b} and Fig.~\ref{fig:fig4e}, the rebuffer ratio and bitrate switches of all algorithms increased quite a bit in comparison to the WiFi environment.
MPC adopts a rule-based method to predict future bandwidth based on past throughput, which is incapable of capturing the complex evolutionary trends in the network.
Fugu incorporates a neural network for prediction; however, its simple network structure and insufficient use of essential information result in low accuracy.
As a buffer-based scheme, the increasing difficulty of throughput prediction has little influence on the performance of BBA.
Nevertheless, the pronounced fluctuation of 4G networks makes it difficult for the reservoir to absorb the variation, which also leads to a higher rebuf ratio.

Remarkably, RobustMPC demonstrates exceptional adaptability to volatile networks.
While its conservative strategy leads to some quality loss, the overall QoE score still improves due to the reduced rebuffer ratio.
As a wastage-aware variant of RobustMPC, PSWA also attains relatively high QoE.
However, its aggressive downloading strategy results in a certain degree of QoE decline, making it inferior to BE-ABR.
Our Transformer-based $\mathrm{T^{3}P}$ method fully leverages the potential of the Transformer architecture to exploit the underlying mechanisms and patterns of the networks.
Consequently, it achieves a lower rebuffer ratio than the other algorithms and is nearly on par with the QoE performance of conservative RobustMPC on highly volatile 4G networks.

\textcolor{blue2}{We further analyze two concrete playback cases in WiFi environments, providing a detailed showcase of BE-ABR's bitrate selections and buffer dynamics in Fig.~\ref{fig:fig70}.
The first case illustrates BE-ABR's behavior under a period of low network bandwidth, while the second depicts its performance during a period where the available network bandwidth exceeds the video's highest bitrate configuration.
BE-ABR demonstrates effective buffer management under both network conditions. 
Notably, in the high-bandwidth scenario, the buffer volume does not continuously increase; instead, it stabilizes at a relatively low level.}

\textcolor{blue2}{In our real-world evaluations, we conduct tests in school WiFi environments with limited bandwidth to ensure that the bandwidth would not consistently exceed the video's highest bitrate.
Moreover, we also conduct small-scale experiments using high-speed home WiFi with an average bandwidth of 68.13Mbps.
In this environment, different ABR algorithms demonstrate similar QoE performance, yet BE-ABR significantly reduces data wastage by 68.80\%$\sim$88.93\%.}

\begin{figure*}[!t]
	\centering
        \captionsetup[subfloat]{font=footnotesize}
	\subfloat[4G/LTE \& $f_1$ mode]{\includegraphics[height=1in]{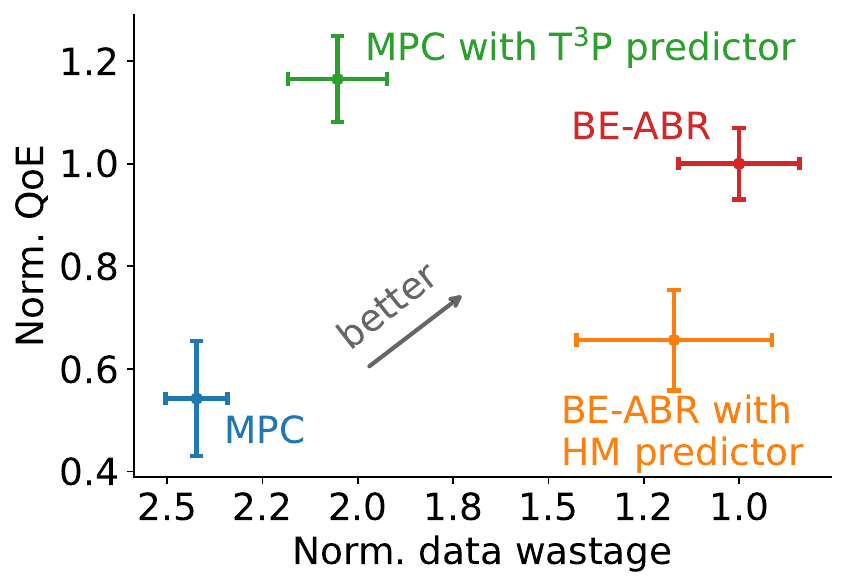}%
		\label{fig:fig5a}}
	\hfil
  	\subfloat[4G/LTE \& $f_2$ mode]{\includegraphics[height=1in]{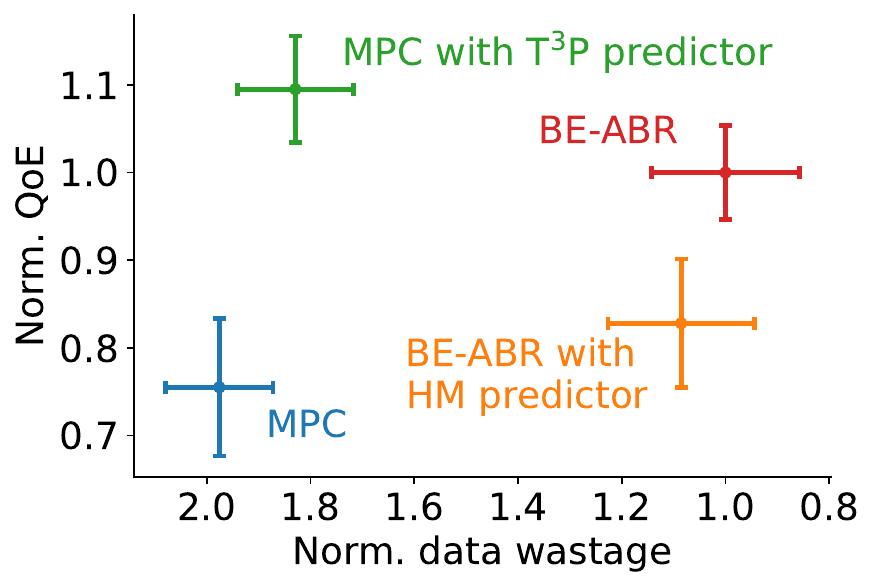}%
		\label{fig:fig5b}}
	\hfil
	\subfloat[Quality]{\includegraphics[height=1in]{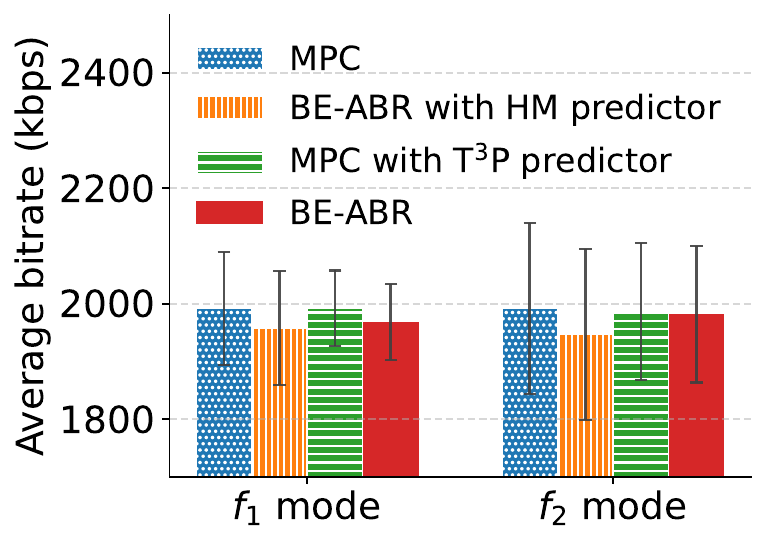}%
		\label{fig:fig5c}}
        \hfil
        \subfloat[Rebuffer ratio]{\includegraphics[height=1in]{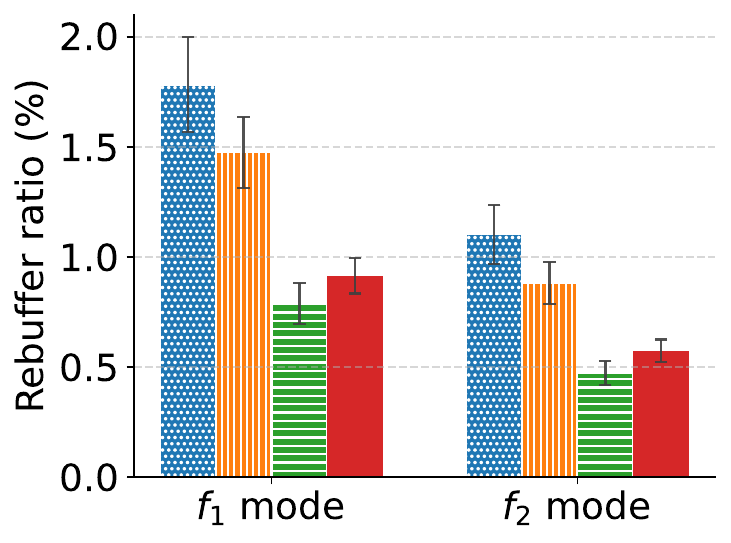}%
		\label{fig:fig5d}}
        \hfil
        \subfloat[Switch]{\includegraphics[height=1in]{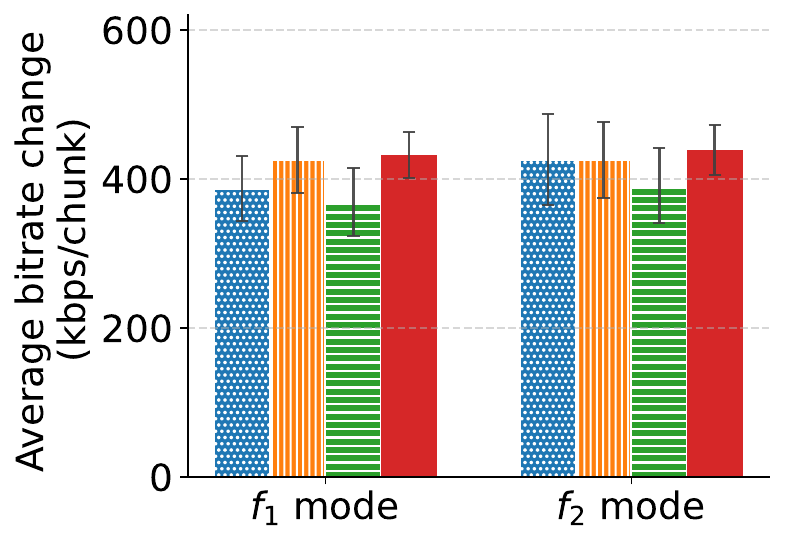}%
		\label{fig:fig5e}}
	\hfil
        \caption{Results from the component-wise study analysis. Panels (a) and (b) showcase component-wise contributions on QoE performance and traffic wastage under two user behavior modes over real 4G/LTE networks. Metrics are normalized against the performance of BE-ABR. \textcolor{myblue}{QoE metrics are computed as $QoE_{lin}$. Panels (c), (d), and (e) further showcase the performance of three QoE-related assessment indexes.}}
	\label{fig:fig5}
\end{figure*}

\subsection{Component-Wise Study}
In this study, we systematically replace key components of BE-ABR to better understand their contributions to the end-to-end system performance.
BE-ABR comprises two main modules: a rate adaptation controller with active, fine-grained buffer management and a Transformer-based, time-aware transmission delay predictor ($\mathrm{T^{3}P}$).
To verify the effects of each module, we keep either the predictor or the controller constant, while replacing the other with the component from the MPC \cite{yin2015control} system.

In Fig.~\ref{fig:fig5}, we compare the system performance of the original MPC, MPC's controller combined with $\mathrm{T^{3}P}$ predictor, BE-ABR's controller combined with HM predictor and original BE-ABR under 4G/LTE networks with varying user behavior assumptions.
Fig.~\ref{fig:fig5c}, Fig.~\ref{fig:fig5d}, and Fig.~\ref{fig:fig5e} further illustrates the individual terms considered in the QoE computation.

First, compared to the original MPC, substituting the predictor with $\mathrm{T^{3}P}$ significantly enhances QoE performance, yielding an increase of 1.45-2.15$\times$.
As shown in Fig.~\ref{fig:fig5d}, a considerable portion of the QoE improvement is attributed to $\mathrm{T^{3}P}$'s capability to accurately predict network fluctuations and avoid rebuffering.
In particular, the rebuffer ratio is reduced by 56.6$\%$ on average.
This enhancement demonstrates the contribution and effectiveness of our Transformer-based transmission time prediction design.

Second, retaining the original predictor and incorporating BE-ABR's controller significantly diminishes the traffic waste.
More specifically, it nearly halves the data wastage of the original MPC under any assumed user behavior. 
These results substantiate the resilience and compatibility of BE-ABR's buffer-controlled rate adaptation scheme; even when integrated with a less accurate predictor, it remains effective.
What's more, it is worth noting that replacing the controller does not result in a degradation of QoE performance in comparison to the original MPC.
This is attributed to the practical algorithm designs that we developed to enhance QoE robustness and control QoE loss.

\textcolor{myblue}{
Third, in this component-wise study, the original BE-ABR achieves optimal performance when considering both wastage and QoE from a holistic perspective.
Although the MPC's controller combined with the $\mathrm{T^{3}P}$ predictor exhibits better QoE performance, BE-ABR, at the cost of sacrificing only 8.7\% QoE performance, has achieved a 45.4\% improvement in wastage reduction.}

\begin{figure}[!tbp]
	\centering
	\includegraphics[width=2.8in]{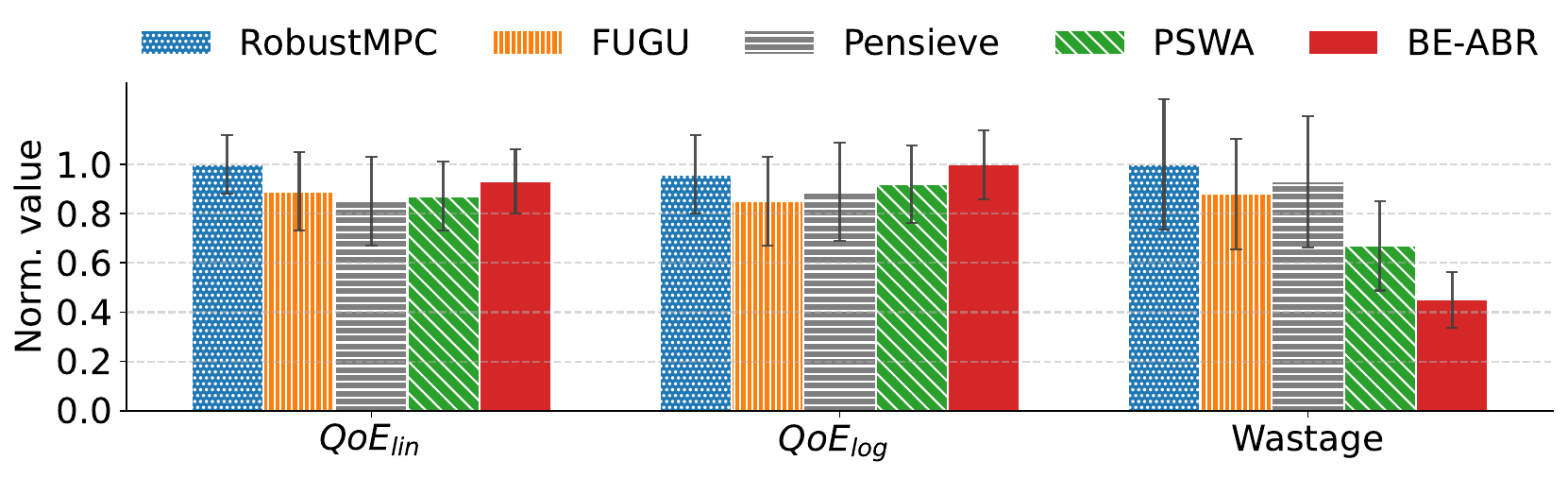}
	\caption{\textcolor{myblue}{Results from the user study analysis. Metrics are normalized against the maximum values. The QoE performance is evaluated using two metrics: $QoE_{lin}$ and $QoE_{log}$.}}
	\label{fig:fig6}
\end{figure}

\textcolor{myblue}{
\subsection{User Study}
Considering that user viewing behavior is influenced by various factors and demonstrates huge heterogeneity, we conduct a user study.
The primary aim of this study was to obtain testing results of algorithm performance under realistic user viewing conditions.
In this experiment, we select four algorithms that have shown top-ranked performance in real-world experiments, which are RobutMPC, FUGU, Pensieve, PSWA, to compare with BE-ABR.}

\textcolor{myblue}{
We invite 40 student participants for this study.
Each participant is shown 15 test videos from the \textit{SD video set}, and they are given the autonomy to exit the current video and switch to the next one at any time based on their preferences.
Participants are randomly assigned to one of the five testing ABR algorithms on an equal basis.
The experiment is hosted in a school classroom, with the network environment same as the public school Wi-Fi used in the real-world experiments.}

\textcolor{myblue}{
In Fig.~\ref{fig:fig6}, we demonstrate the transmission performance of each tested algorithm.
Here, the performance of QoE is gauged with two QoE metrics: $QoE_{lin}$ and $QoE_{log}$.
The results show that BE-ABR's QoE performance is comparable to RobustMPC and superior to the other three algorithms.
BE-ABR also achieves the lowest wastage, reducing waste by 32.84\%-55.10\%.
}

\begin{figure}[!t]
	\centering
        \captionsetup[subfloat]{font=footnotesize}
	\subfloat[$f_1$ mode]{\includegraphics[width=1.72in]{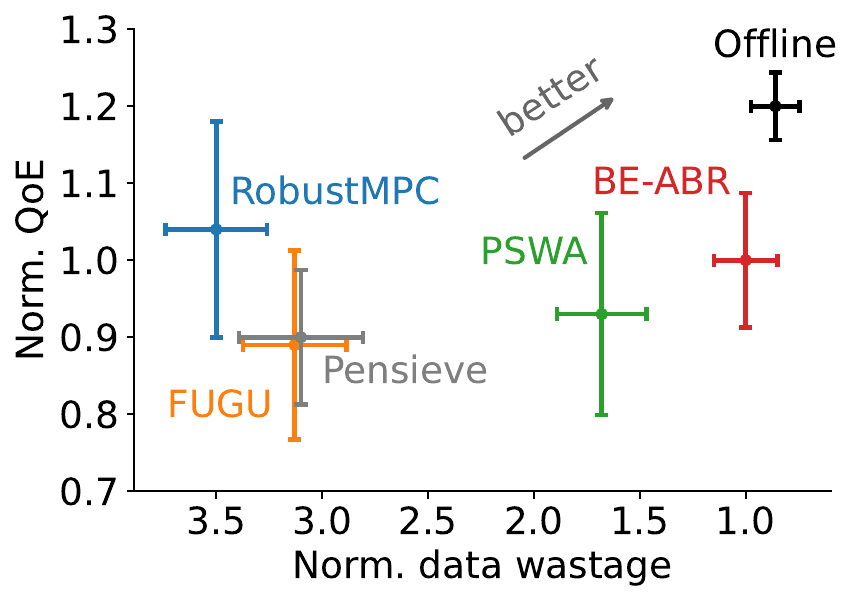}%
		\label{fig:fig7a}}
	\hfil
	\subfloat[$f_2$ mode]{\includegraphics[width=1.72in]{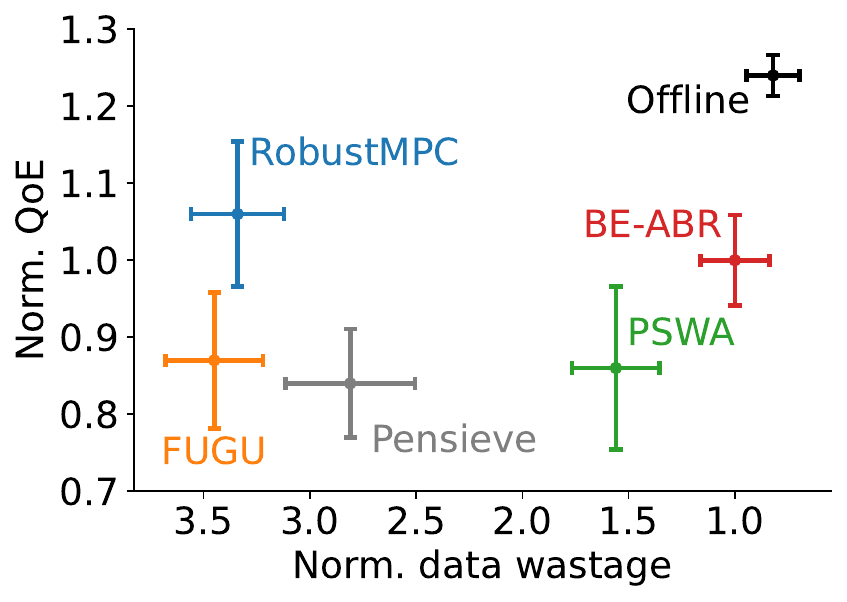}%
		\label{fig:fig7b}}
	\caption{\textcolor{myblue}{Comparison of traffic wastage and QoE performance on Belgium 4G and Lumos 5G datasets. Metrics are normalized against the performance of BE-ABR. QoE metrics are computed as $QoE_{log}$.}}
	\label{fig:fig7}
\end{figure}

\begin{table}[!t]
\footnotesize
\centering
\color{myblue}
\caption{\textcolor{myblue}{Average buffer status during playback.}}
\begin{tabular}{c|ccccc}
\hline
& RobustMPC & Fugu  & Pensieve & PSWA & BE-ABR \\ \hline    
Length\hspace{0.05cm}(s) & 16.61 & 13.28 & 14.42 & 9.15 & 6.29 \\ \hline
Volume\hspace{0.05cm}(MB) & 34.22 & 29.08 & 27.61 & 15.57 & 9.29 \\ \hline
\end{tabular}
\label{bufferState}
\end{table}

\begin{table}[!t]
    \footnotesize
    \centering
    \color{myblue}
    \caption{\textcolor{myblue}{Performance of BE-ABR under different prediction accuracy.}}
    \begin{tabular}{c|c|c|c|c|c|c}
        \hline
        \multirow{2}*{MAPE} & & \multirow{2}*{Ratio} & \multirow{2}*{$QoE_{log}$} & & \multirow{2}*{Ratio} & \multirow{2}*{$QoE_{log}$} \\
        & & & & & & \\ 
        \hline
        \multicolumn{1}{c|}{0$\sim$10\%} & \multirow{3}*{$f_1$} & 0.36 & 1.99 & \multirow{3}*{$f_2$} & 0.34 & 2.07 \\
        \cline{1-1} \cline{3-4} \cline{6-7} 
        10$\sim$20\% & & 0.38 & 1.84 & & 0.33 & 1.94 \\
        \cline{1-1} \cline{3-4} \cline{6-7} 
        \textgreater{}20\% & & 0.26 & 1.58 & & 0.33 & 1.63 \\
        \hline 
    \end{tabular}
    \label{mape}
\end{table}

\textcolor{myblue}{
\subsection{Generalization Ability Test}
This experiment is designed to test the generalization ability of BE-ABR across varying network environments.
Although experiments conducted with real-world network connections greatly restore the real streaming scenarios, their execution complexity is not conducive to multi-scenario testing. 
Therefore, in this experiment, we resort to trace-driven simulation with Mahimahi\cite{netravali2015mahimahi}.}

\textcolor{myblue}{
We employ the widely used Belgium 4G traces~\cite{van2016http} and Lumos 5G traces~\cite{narayanan2020lumos5g}, both of which cover a broad spectrum of mobile scenarios. 
Leveraging these network traces, we perform evaluations on RobustMPC, FUGU, Pensieve, PSWA, and BE-ABR with the \textit{UHD video set}. 
Within each network trace, we run the test videos 10 times for every ABR algorithm under evaluation.}

\textcolor{myblue}{
Figure~\ref{fig:fig7} presents the QoE performance and traffic wastage of each ABR schemes, as well as the theoretical optimum.
BE-ABR exhibits strong environmental stability and achieves optimal overall performance across multiple test scenarios.
Specifically, BE-ABR's QoE performance is second only to RobustMPC, with a decrease of 4.75\%, and it successfully reduced traffic wastage by up to 71.43\%.
Compared to other learning-based methods, BE-ABR improves QoE by 11.11\%-19.05\%. 
Against the wastage-aware PSWA, BE-ABR further reduces wastage by 38.19\%.
Moreover, BE-ABR reaches 81.10\% of the theoretical optimum in QoE and 84.70\% in wastage.
Table~\ref{bufferState} provides a view of the buffer status during playback. 
It is clear to see that BE-ABR maintains a small buffer size during streaming, with the average buffered data volume only 27.15\%-59.67\% of that of other algorithms.}

\textcolor{myblue}{
Additionally, we present the system performance of BE-ABR under various prediction errors in Table~\ref{mape}.
The prediction error is measured using the Mean Absolute Percentage Error (MAPE) metric, which is calculated as:
\begin{equation}
MAPE = \frac{100\%}{n} \times \sum_{i=1}^{n} \left| \frac{\hat{y}_i - y_i}{y_i} \right|
\end{equation}
Here, $y_i$ signifies the actual value, while $\hat{y}_i$ represents the predicted value.
As revealed by the data in the table, there is a direct correlation between the accuracy of the prediction and the QoE performance—the higher the prediction accuracy, the more improved the QoE performance. 
When the prediction error surpasses 20\%, there is a corresponding 21.26\% decrease in QoE. 
However, our T3P algorithm effectively mitigates this, with 70.5\% of prediction errors falling below 20\%.}

\subsection{Evaluations for $T^{3}P$}\label{subsec:EVA_TTP}
The component-wise analysis conducted in the preceding section demonstrated that $\mathrm{T^{3}P}$ considerably enhances the overall system performance. 
This finding is exciting, especially as network prediction continues to be a focal point of research in video transmission systems.
Therefore, in this section, we further evaluate the predictive capabilities of $\mathrm{T^{3}P}$, comparing it with several network predictor baselines using large-scale, real-world network datasets.

\textbf{Algorithms for comparison:}
Network prediction is a fundamental and significant subproblem in video streaming systems, and there has been extensive research dedicated to its study.
we select four of the most typical and representative network predictor baselines for comparison.
\textcolor{myblue}{
We carry out a 1-to-1 replication of the comparison algorithms, using the model structure parameters recommended in their papers (or presented in the public code).} 
These comparison algorithms include: 
\begin{itemize}
    \item \textit{HM:}
    The HM method is first adopted in MPC\cite{yin2015control}, and it has since become widely used in video streaming systems due to its simplicity and ease of deployment.
    It predicts future bandwidth by calculating the harmonic mean value of several past throughput samples, and estimates transmission time by taking the ratio of chunk sizes to the predicted bandwidth.
    \item \textit{SVM:}
    The Support Vector Machine (SVM) is a relatively simple supervised machine learning algorithm.
    Raca \textit{et al.}\cite{raca2020leveraging} utilize the SVM model to predict the achievable throughput in cellular networks using wireless network parameters.
    \item \textit{LSTM:}
    Long Short-Term Memory (LSTM) is a type of Recurrent Neural Network (RNN) that is particularly adept at learning long-term dependencies. 
    Mei \textit{et al.}\cite{mei2019realtime} propose using an LSTM model to predict real-time mobile bandwidth based on historical observed throughput.
    \item \textit{MLP:}
    Yan \textit{et al.}\cite{yan2020learning} train a Multilayer Perceptron (MLP) to give direct prediction of transmission time.
    The model takes historical chunk information and network statistics as inputs and outputs a discretized probability distribution of predicted transmission times.
\end{itemize}

\textbf{Datasets:}
We download a large-scale real-world network dataset from Puffer to generate the testing dataset.
Specifically, we gather 10 million samples over a 20-day period to ensure the dataset's generality and diversity.
As a video-streaming website open to the public, the recorded data in Puffer contains various network scenarios, such as 4G, 5G, and WiFi, as well as different network states, including stable and fluctuating. 
Therefore, our testing dataset can effectively reflect the characteristics of different network environments and serve as an eligible test set for evaluating prediction algorithms' performance.

\begin{table}[!tbp]
    \footnotesize
	\caption{Comparison over multiple error-based evaluation metrics.}
        \arrayrulecolor{black}
	\centering
	\begin{tabular}{ccccc}
		\toprule
		& MAE $\downarrow$ & RMSE $\downarrow$ & MAPE $\downarrow$ \\
		\midrule
		HM & 0.538 & 0.943 & 40.2\% \\
		SVM & 0.402 & 0.651 & 35.7\% \\
		LSTM & 0.338 & 0.378  & 30.6\% \\
		MLP & 0.246 & 0.316 & 26.1\%\\
		\textbf{$\rm T^{3}P$} & \textbf{0.178} & \textbf{0.262}  & \textbf{16.6\%}\\
		\bottomrule
	\end{tabular}
\label{tab:t3p}
\end{table}

\begin{figure}[!tbp]
	\centering
	\includegraphics[width=2.8in]{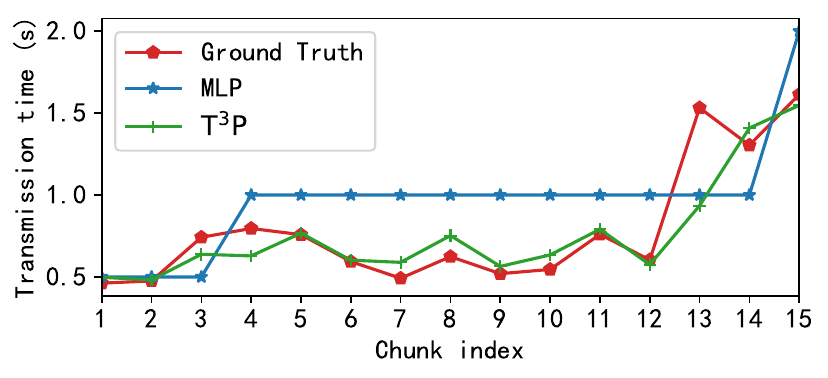}
	\caption{A case study showcasing prediction results of $\mathrm{T^{3}P}$ and MLP.}
	\label{fig:fig8}
\end{figure}

\textbf{Predictive performance over real-world networks:}
We evaluate the performance of prediction algorithms using three well-established metrics, including mean absolute error (MAE), root mean squared error (RMSE), and mean absolute percentage error (MAPE).
The comprehensive results are presented in Table~\ref{tab:t3p}, where $\mathrm{T^{3}P}$ model demonstrates superior performance in all metrics compared to state-of-the-art prediction algorithms. 
Specifically, our $\mathrm{T^{3}P}$ model reduces the MAE, RMSE and MAPE by 27.6-66.9\%, 17.1-72.2\% and 36.4-58.7\%, respectively.

Rule-based methods such as HM \cite{yin2015control} and SVM \cite{raca2020leveraging} are not capable of learning the complex relationship among variables in the network environment.
LSTM\cite{mei2019realtime} applies a deep learning technique for network predicting.
However, only relying on historical throughput information is not enough for accurate prediction, as available bandwidth is affected by a multitude of factors from both lower and upper layers of the communication stack.
MLP\cite{yan2020learning} develops a neural network and consider a variety of noisy inputs, which made certain achievements.
Nevertheless, it neglected the influence of irregular sampling time and employs a simplistic network structure.
In contrast, our $\mathrm{T^{3}P}$ model fully leverages the potential of the Transformer architecture and incorporates three functional modules tailored to network characteristics, resulting in a significant enhancement of predictive ability. 
\textcolor{myblue}{To showcase the accuracy of the $\mathrm{T^{3}P}$ prediction results, we have included a case illustration in Fig.~\ref{fig:fig8}.}

\begin{figure}[!tbp]
	\centering
	\includegraphics[width=2.3in]{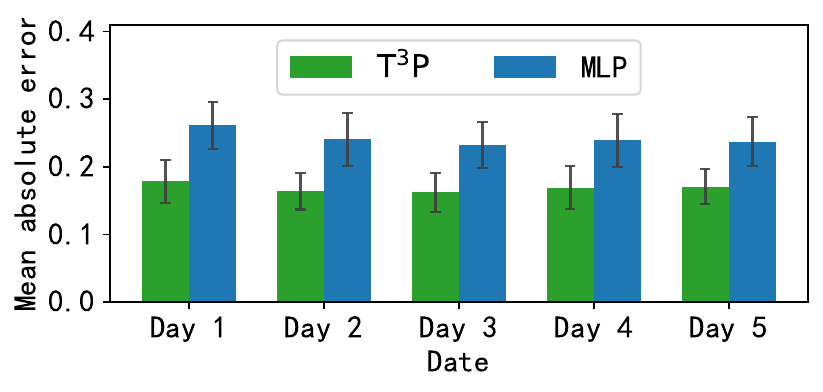}
	\caption{Variation in the Mean Absolute Error (MAE) of different transmission time prediction algorithms across consecutive days.}
	\label{fig:fig9}
\end{figure}

\textbf{Temporal performance variation:}
Fig.~\ref{fig:fig9} illustrates the variations in prediction accuracy for consecutive days of $\mathrm{T^{3}P}$ and MLP algorithms.
It is observed that the performance of both algorithms exhibits minor fluctuations across different dates.
This variation is expected, as the network prediction problem becomes more challenging during specific periods due to the dynamic and random network state changes in real-world scenarios.
For instance, predicting network conditions during weekends can be more difficult due to increased network congestion and higher traffic volumes. 
Remarkably, our proposed algorithm demonstrates robust performance across multiple days, consistently achieving low-error predictions.

\subsection{System Overhead}
Examining system overhead to understand resource consumption is essential for evaluating system performance. 
High overhead could negatively impact the overall performance and response speed of the system, potentially leading to performance bottlenecks.
In this section, we examine the main overheads of BE-ABR system.

\textbf{Model training of $\mathrm{T^{3}P}$:}
The training process of the $\mathrm{T^{3}P}$ model is conducted offline, with the trained parameters subsequently loaded for users.
An Ubuntu 20.04 server, equipped with 64 Intel Xeon Gold 6130 processors, 8 GeForce RTX 2080 Ti GPUs, and 512 GB of RAM, is utilized to train the $\mathrm{T^{3}P}$ model on a vast real-world network dataset obtained from Puffer.
Based on our measurements, the training resource occupancy amounts to 1045 MB of GPU memory, and the average time for the model to reach convergence is approximately 50 minutes, indicating that the training of $\mathrm{T^{3}P}$ has an acceptable overhead.

\textbf{Inference latency of $\mathrm{T^{3}P}$:}
Typically, real-time inference of the $\mathrm{T^{3}P}$ model is carried out on the resource-limited client end, which is emulated using a Mini PC (\ref{subsec:exp_setup}) in our experiment.
As an online algorithm, the inference latency of $\mathrm{T^{3}P}$ must be sufficiently short to avoid adversely impacting the system's overall performance.
Our measurements show an average inference latency of 8.8 ms for the $\mathrm{T^{3}P}$ model, a figure that aligns with the requirements for preserving system efficiency.

\textbf{Optimal configurations searching:}
To expedite the search time for the optimal configuration, we introduce the Genetic Algorithm, which is elaborately described in \ref{subsec:abr}.
According to our gathered data, the GA typically finds the solution within an average timeframe of 76 ms, \textcolor{myblue}{reaching a remarkable 96.3\% efficiency compared to the optimal solution's performance.}
This is a significantly reduced search time, especially when compared to exhaustive search methods. 
In the context of a control system that operates at the second level, this time frame is deemed quite acceptable, ensuring real-time responsiveness and high system performance.

\section{Limitation and Future work}
Despite the promising findings, this work still has room for improvement and presents several opportunities for further investigation.
First, although the proposed method demonstrates exceptional performance in WiFi networks, enhancing its adaptability to the highly variable and uncertain 4G/LTE networks demands more in-depth examinations. 

Second, our current findings reveal that BE-ABR requires an average of 86 ms for a single inference.
For a video transmission system operating at a control granularity of seconds, this inference latency is considered to be acceptable.
But in the future, we will focus on researching lightweighting techniques to reduce the computational complexity and achieve faster response times, enabling the adaptation to streaming systems with finer control granularity.

Third, traffic wastage is a prevalent concern across various video systems, encompassing live streaming, short video streaming and so on.
Investigating the potential extension of the concepts presented in this paper to other video systems constitutes a worthwhile direction for future research.

\section{Conclusion}
In this research, we design and implement a bandwidth-efficient bitrate adaptation scheme (called BE-ABR) for video streaming systems, concurrently optimizing both QoE enhancement and bandwidth resource conservation.
Our work commences with the formulation of buffered data volume dynamics, which lays the foundation for the design of BE-ABR.
The proposed BE-ABR algorithm comprises a Transformer-based time-aware transmission delay predictor ($\mathrm{T^{3}P}$) and a rate adaptation algorithm with fine-grained buffer control.
Comprehensive experiments conducted on real-world wireless networks demonstrate that the overall performance of BE-ABR significantly exceeds existing approaches.

\ifCLASSOPTIONcaptionsoff
  \newpage
\fi

\bibliographystyle{IEEEtran}
\bibliography{IEEE}

\begin{IEEEbiography}[{\includegraphics[width=1in,height=1.25in,clip,keepaspectratio]{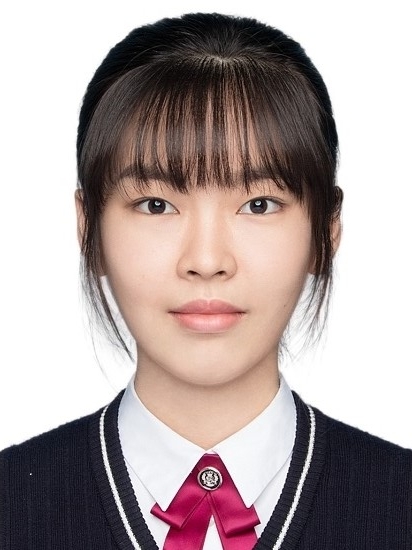}}]{Hairong Su}
  received her B.Sc. degree from Xi'an Jiaotong University (XJTU), China in 2021. She is currently working toward a Ph.D. degree in the National Engineering Laboratory for Big Data Analytics at XJTU. In 2020, she has been a visiting student with the Department of Mathematics, Georgia Institute of Technology. Her research interests include video streaming, networked systems and deep learning.
\end{IEEEbiography}
\vspace{-7 mm}
\begin{IEEEbiography}[{\includegraphics[width=1in,height=1.25in,clip,keepaspectratio]{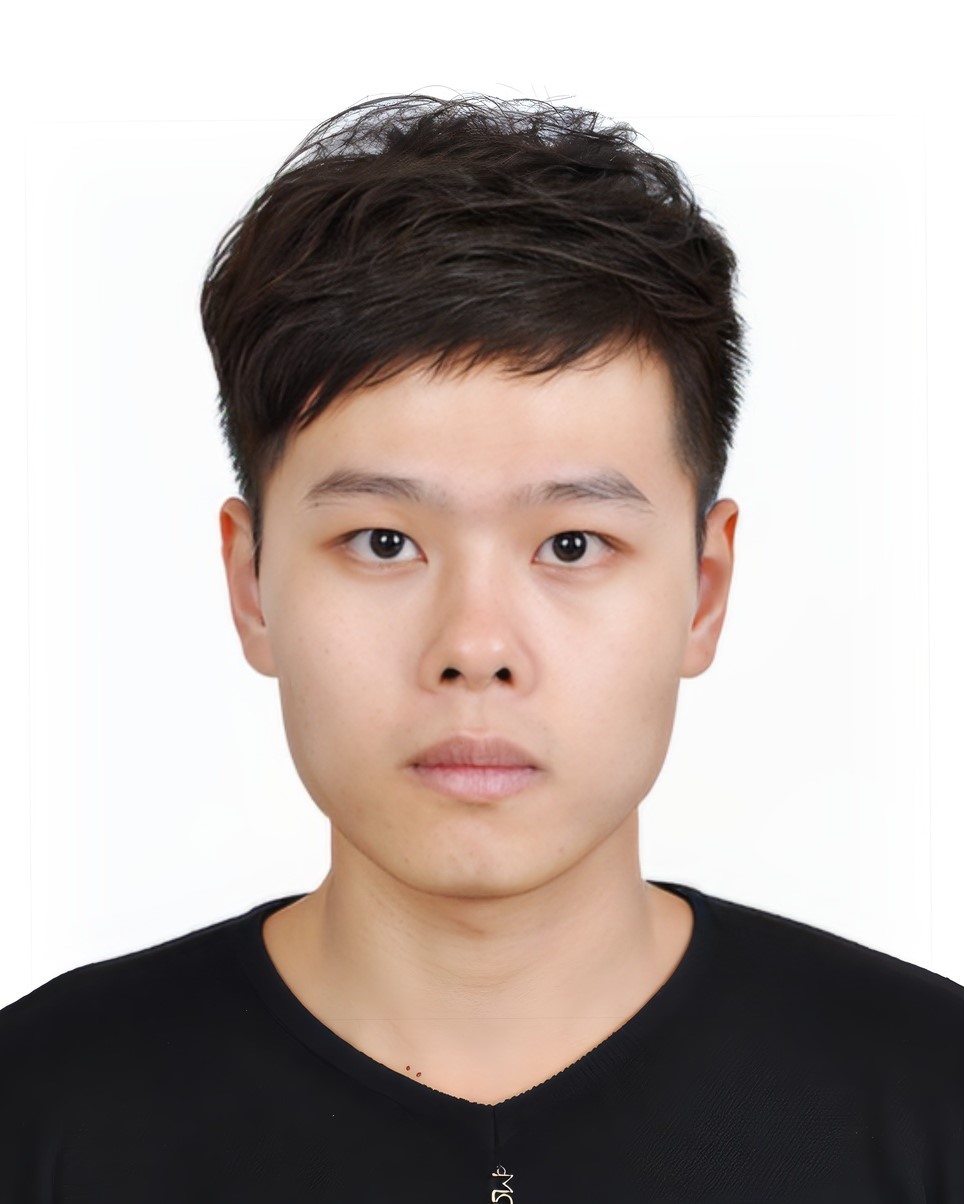}}]{Shibo Wang}
received his B.Sc. degree from Xi’an Jiaotong University (XJTU), China in 2019. He is currently working toward a Ph.D. degree in the
National Engineering Laboratory for Big Data Analytics at XJTU. He is also currently working as a research assistant at the Department of Computer Science and Engineering at The Chinese University of Hong Kong (CUHK), China. His research interests include low-latency networking systems, congestion control, real-time communication, 360-degree video streaming, and video analytics.
\end{IEEEbiography}
\vspace{-7 mm}
\begin{IEEEbiography}[{\includegraphics[width=1in,height=1.25in,clip,keepaspectratio]{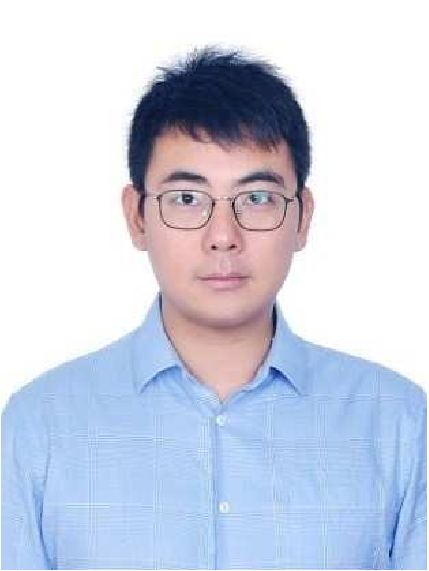}}]{Shusen Yang}
  received his Ph.D. degree in Computing from Imperial College London in 2014. He is a professor and director of the National Engineering Laboratory for Big Data Analytics, and deputy director of Ministry of Education (MoE) Key Lab for Intelligent Networks and Network Security, both at Xi’an Jiaotong University (XJTU), China. Before joining XJTU, he worked as a lecturer (assistant professor) at the University of Liverpool from 2015 to 2016, and a research associate at Intel Collaborative Research Institute (ICRI) from 2013 to 2014. He is a DAMO Academy Young Fellow, and an honorary research fellow at Imperial College London. He is a senior member of IEEE and a member of ACM. His research focuses on distributed systems and data sciences, and their applications in industrial scenarios.
\end{IEEEbiography}
\vspace{-7 mm}
\begin{IEEEbiography}[{\includegraphics[width=1in,height=1.25in,clip,keepaspectratio]{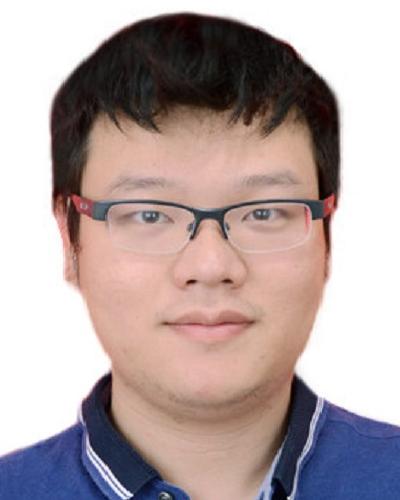}}]{Tianchi Huang}
received his Ph.D. degree in the Department of Computer Science and Technology at Tsinghua University in 2023. His research work focuses on multimedia network streaming, including transmitting streams, and edge-assisted content delivery. He has been the reviewer for IEEE TRANSACTIONS ON VEHICULAR TECHNOLOGY, IEEE TRANSACTIONS ON MOBILE COMPUTING, and IEEE TRANSACTIONS ON MULTIMEDIA. He received the Best Student Paper Award presented by ACM Multimedia System 2019 Workshop and Best Paper Nomination presented by ACM Multimedia Asia 2022. He currently works in the R\&D department at Sony.
\end{IEEEbiography}
\vspace{-7 mm}
\begin{IEEEbiography}[{\includegraphics[width=1in,height=1.25in,clip,keepaspectratio]{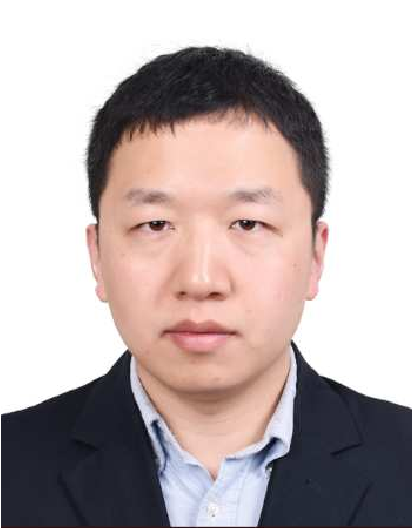}}]{Xuebin Ren}
   (Member, IEEE) received the Ph.D. degree from the Department of Computer Science and Technology, Xi’an Jiaotong University, China, in 2017. He is currently an Associate Professor with Xi’an Jiaotong University and a member of the National Engineering Laboratory for Big Data Analytics (NEL-BDA). He has been a Visiting Ph.D. student with the Department of Computing, Imperial College London, from 2016 to 2017. His research interests include data privacy protection, federated learning, and privacy-preserving machine learning. He is a member of the ACM.
\end{IEEEbiography}

\end{document}